\documentclass{aa}

\usepackage{graphicx}
\usepackage{txfonts}
\usepackage{natbib}     
\usepackage{xcolor} 
\usepackage{threeparttable} 
\usepackage[colorlinks=true,citecolor=blue]{hyperref} 

\usepackage[version=4]{mhchem}

\bibpunct{(}{)}{;}{a}{}{,} 

\begin{document}

   \title{LBT transmission spectroscopy of HAT-P-12b }
   
    \subtitle{confirmation of a cloudy atmosphere with no significant alkali features}

   \author{F. Yan\inst{1,2}
          \and
          N. Espinoza\inst{3}
          \and
          K. Molaverdikhani\inst{2}
          \and
                  Th. Henning\inst{2}
          \and
          L. Mancini\inst{2,4,5,6}
          \and
          M. Mallonn\inst{7}
          \and
          B. V. Rackham\inst{8,9,10}\thanks{51 Pegasi b Fellow}
          \and
                  D. Apai\inst{2,8,11}
                  \and
                  A. Jord\'an\inst{12,13,2}
          \and
          P. Molli\`ere\inst{2}
          \and
          G. Chen\inst{14}
          \and
          L. Carone\inst{2}
          \and
          A. Reiners\inst{1}
                }

   \institute{Institut f\"ur Astrophysik, Georg-August-Universit\"at, Friedrich-Hund-Platz 1, 37077 G\"ottingen, Germany\\
\email{fei.yan@uni-goettingen.de}
                                \and
                                    Max Planck Institute for Astronomy, K\"onigstuhl 17, 69117, Heidelberg, Germany
                                 \and
                                 Space Telescope Science Institute, 3700 San Martin Drive, Baltimore, MD 21218, USA
                                 \and
                                 Department of Physics, University of Rome Tor Vergata, Via della Ricerca Scientifica 1, I-00133 Rome, Italy
                                 \and
                                 INAF – Osservatorio Astrofisico di Torino, via Osservatorio 20, I-10025 Pino Torinese, Italy
                                 \and
                                 International Institute for Advanced Scientific Studies (IIASS), Via G. Pellegrino 19, I-84019 Vietri sul Mare (SA), Italy
                                 \and Leibniz-Institut f\"ur Astrophysik Potsdam (AIP), An der Sternwarte 16, 14482 Potsdam, Germany
                                 \and Department of Astronomy/Steward Observatory, The University of Arizona, 933 N. Cherry Avenue, Tucson, AZ 85721, USA
                                \and Department of Earth, Atmospheric and Planetary Sciences, Massachusetts Institute of Technology, 77 Massachusetts Avenue, Cambridge, MA 02139, USA                                 
                                \and Kavli Institute for Astrophysics and Space Research, Massachusetts Institute of Technology, 77 Massachusetts Avenue, Cambridge, MA 02139, USA                                 
                                 \and Lunar and Planetary Laboratory, The University of Arizona, 1640 E. Univ. Blvd. Tucson, AZ 85721, USA
                                 \and Facultad de Ingenier\'ia y Ciencias, Universidad Adolfo Ib\'a\~nez, Av.\ Diagonal las Torres 2640, Pe\~nalol\'en, Santiago, Chile
                                 \and Millennium Institute of Astrophysics, Av. Vicu na Mackenna 4860, 782-0436 Macul, Santiago, Chile
                                \and Key Laboratory of Planetary Sciences, Purple Mountain Observatory, Chinese Academy of Sciences, Nanjing 210023, PR China
                                \\     
         }
        \date{Received December 05, 2019; accepted July 30, 2020}


  \abstract
 {The hot sub-Saturn-mass exoplanet HAT-P-12b is an ideal target for transmission spectroscopy because of its inflated radius. We observed one transit of the planet with the  multi-object double spectrograph (MODS) on the Large Binocular Telescope (LBT) with the binocular mode and obtained an atmosphere transmission spectrum with a wavelength coverage of $\sim$ 0.4 -- 0.9 $\mathrm{\mu}$m. The spectrum is relatively flat and does not show any significant sodium or potassium absorption features. Our result is consistent with the revised \textit{Hubble Space Telescope} (\textit{HST}) transmission spectrum of a previous work, except that the \textit{HST} result indicates a tentative detection of potassium. The potassium discrepancy could be the result of statistical fluctuation of the \textit{HST} dataset. 
We fit the planetary transmission spectrum with an extensive grid of cloudy models and confirm the presence of high-altitude clouds in the planetary atmosphere. The fit was performed on the combined LBT and \textit{HST} spectrum, which has an overall wavelength range of 0.4 -- 1.6 $\mathrm{\mu}$m. 
The LBT/MODS spectrograph has unique advantages in transmission spectroscopy observations because it can cover a wide wavelength range with a single exposure and acquire two sets of independent spectra simultaneously.
 }

   \maketitle

%

\section{Introduction}
Transmission spectroscopy is one of the main techniques for exoplanet atmosphere characterizations. A transmission spectrum provides valuable information of the planet, such as the atmospheric composition and structure.
The first detection of an exoplanet atmosphere \citep{Charbonneau2002} revealed atomic sodium (Na) in HD 209458b. 
Other species, including atomic potassium (K), $\mathrm{H_2O}$, $\mathrm{CO}$, hydrogen, and helium, have also been detected in exoplanet atmospheres \citep[e.g.,][]{Sing2011, Brogi2016, Kreidberg2018, Yan2018, Nortmann2018, Spake2018}.

In addition to these atoms and molecules, clouds or hazes can also exist in exoplanet atmospheres. They produce a flat transmission spectrum or a nongray scattering slope. The presence of clouds has been proven to be common \citep[e.g.,][]{Sing2011b, Jordan2013, Kreidberg2014}, and it weakens or even mutes the atomic or molecular features. Thus the characterization of clouds is a key challenge in the field of exoplanet atmospheres, although both the formation and nature of these clouds are still unclear. 
The scattering slope can be used to constrain cloud properties, such as particle sizes, and the combination of a continuum spectrum and the strength of atomic and molecular features constrains the atmospheric conditions. For example, \cite{Heng2016} proposed a method for constraining atmospheric cloudiness using Na/K absorption, and \cite{Stevenson2016} explored the use of $\mathrm{H_2O}$ absorption and J-band continuum to constrain atmospheric cloudiness.

A large number of exoplanets has been observed with the \textit{Hubble Space Telescope} (\textit{HST}) \citep[e.g.,][hereafter S2016]{Sing2016}. The \textit{HST} observations have special advantages because they are not affected by the telluric atmosphere, especially in the near-infrared wavelengths where $\mathrm{H_2O}$ and CO absorption features can be present in the atmospheres of Earth and the targeted exoplanet.
Ground-based observations have also proved to be successful in correcting for the telluric effects by applying the differential spectrophotometry method. These observations are mostly carried out with multi-object spectrographs mounted on large telescopes, for example, FORS2 on the Very Large Telescope (VLT) \citep{Lendl2016, Nikolov2016, Sedaghati2016}, OSIRIS on the Gran Telescopio Canarias \citep{Chen2017, Murgas2017}, GMOS on the Gemini telescope \citep{Gibson2013,Todorov2019}, IMACS on the Magellan telescope \citep{Rackham2017, Espinoza2019}, and MODS on the Large Binocular Telescope (LBT) \citep{Mallonn2016}. These ground-based observations provide transmission spectra in the optical wavelength range that cover features such as Na/K \citep{Sing2012, Lendl2017, Chen2018, Nikolov2018, Pearson2019}, TiO/VO \citep{Sedaghati2017}, and the Rayleigh-scattering slopes \citep{Nikolov2015,Parviainen2016,Kirk2017,Chen2017GJ3470}.

\object{HAT-P-12b} is a mildly irradiated planet with an equilibrium temperature of 960 K. The planet has a very low density with a radius of 0.92 $R_\mathrm{J}$ and a mass of 0.20 $M_\mathrm{J}$ \citep{Mancini2018}, thus it is a good target for transmission spectroscopy.
\cite{Line2013} and \cite{Tsiaras2018} analyzed the near-infrared transit data of the planet (1.1 to 1.6 $\mathrm{\mu}$m) taken with the Wide Field Camera 3 (WFC3) of \textit{HST} and found that the planet lacks a cloud-free deep $\mathrm{H_2O}$ absorption feature, suggesting there are high-altitude clouds in its atmosphere. \cite{Sing2016} obtained the optical transmission spectrum of the planet using the Space Telescope Imaging Spectrograph (STIS) mounted on \textit{HST} and discovered a Rayleigh-scattering feature as well as indications for K absorption. \cite{Deibert2019} reported a tentative detection of Na with the high-dispersion spectrograph mounted on the Subaru telescope, but they were not able to detect the K feature.

\cite{Alexoudi2018} (hereafter A2018) reanalyzed the optical \textit{HST} data to investigate a discrepancy of the S2016 result compared to a previously published ground-based broadband transmission spectrum of HAT-P-12b, which did not show a slope \citep{Mallonn2015}. With updated planetary parameters from ground-based photometry, A2018 derived a flat transmission spectrum from the HST data without a significant scattering slope. They concluded that using inaccurate planetary parameters (e.g., orbital inclination) might result in a Rayleigh-scattering-like slope.

In this work, we present a ground-based transit observation of HAT-P-12b with LBT. The obtained transmission spectrum is relatively flat with no significant alkali features.
The paper is organized as follows. In Section 2 we describe the observation and data reduction procedures. In Section 3 we describe the transit light-curve analysis method. In Section 4 we present the obtained transmission spectrum with discussions. Conclusions are presented in Section 5.

\section{Observation and data reduction}

\subsection{LBT observation}
We observed a full transit of HAT-P-12b on 25 March 2017  with the Multi-Object Double Spectrograph \citep[MODS,][]{Pogge2010} mounted on LBT. MODS is a pair of low- to medium-resolution Multi-Object Double CCD spectrographs and imagers with a field of view of $6'\times6'$.  We used the dual-channel mode of LBT, which employs a dichroic to split the incoming beam into separate red- and blue-optimized spectrograph channels at $\sim$ 5600 \AA. The spectrograph has an overall wavelength range of 3200 to 10,000 \AA. We used the G400L grating for the blue channel (3200 \AA -- 5600 \AA) and the G470L grating for the red channel (5600 \AA -- 10,000 \AA). The resolving power of the gratings is $\sim$ 2000 when a narrow slit is used. We used a custom multi-object-spectroscopy (MOS) mask to simultaneously observe a comparison star and the target star. The MOS mask is composed of two wide slits with a width of $10''$ to minimize flux loss and a length of $30''$ to allow sky background subtraction. One of the slits was placed on the target (HAT-P-12), and the other slit was placed on the comparison star (GSC2 N130301284). The color and brightness of the comparison star are similar to those of the target star. According to the USNO-B catalog \citep{Monet2003}, their magnitude differences are $\Delta B = 0.30 $, $\Delta R =  0.41 $, and $\Delta I = 0.39 $.

MODS is a pair of instruments (MODS1 and MODS2) that are individually attached on each of the twin LBT mirrors. We used the binocular mode for the observations and obtained two independent data sets from MODS1 and MODS2. In the rest of the paper, we assign MODS1-B for the blue channel and MODS1-R for the red channel of the MODS1 instrument, and MODS2-B and MODS2-R for the two channels of the MODS2 instrument.

The observations were continuously performed from 07:30 UT to 11:31 UT with 60 s exposure times. The night was photometric, and seeing varied between $0.8'' - 1.3''$. The measured full width at half maximum (FWHM) over the course of the observation is plotted in Fig.~\ref{FWHM}.
The $2\times2$ pixel binning mode was used to reduce the CCD readout time. During the four-hour-long observations, we obtained 125, 107, 127, and 132 spectral frames from MODS1-B, MODS1-R, MODS2-B, and MODS2-R, respectively.

\subsection{Data reduction}
The data reduction was performed with custom IDL scripts. The bias and flat calibrations were performed using the median of 30 bias frames and 30 flatfield frames. Cosmic rays were removed using the \texttt{L.A.Cosmic} tool \citep{Dokkum2001}\footnote{http://www.astro.yale.edu/dokkum/lacosmic/}. We used the lamp spectra (HgAr, XeKr, and Ne lamps) taken with the MOS mask for wavelength calibrations, and the calibration accuracy is $\sim$ 0.4 \AA. 

\begin{figure}
   \centering
   \includegraphics[width=0.4\textwidth]{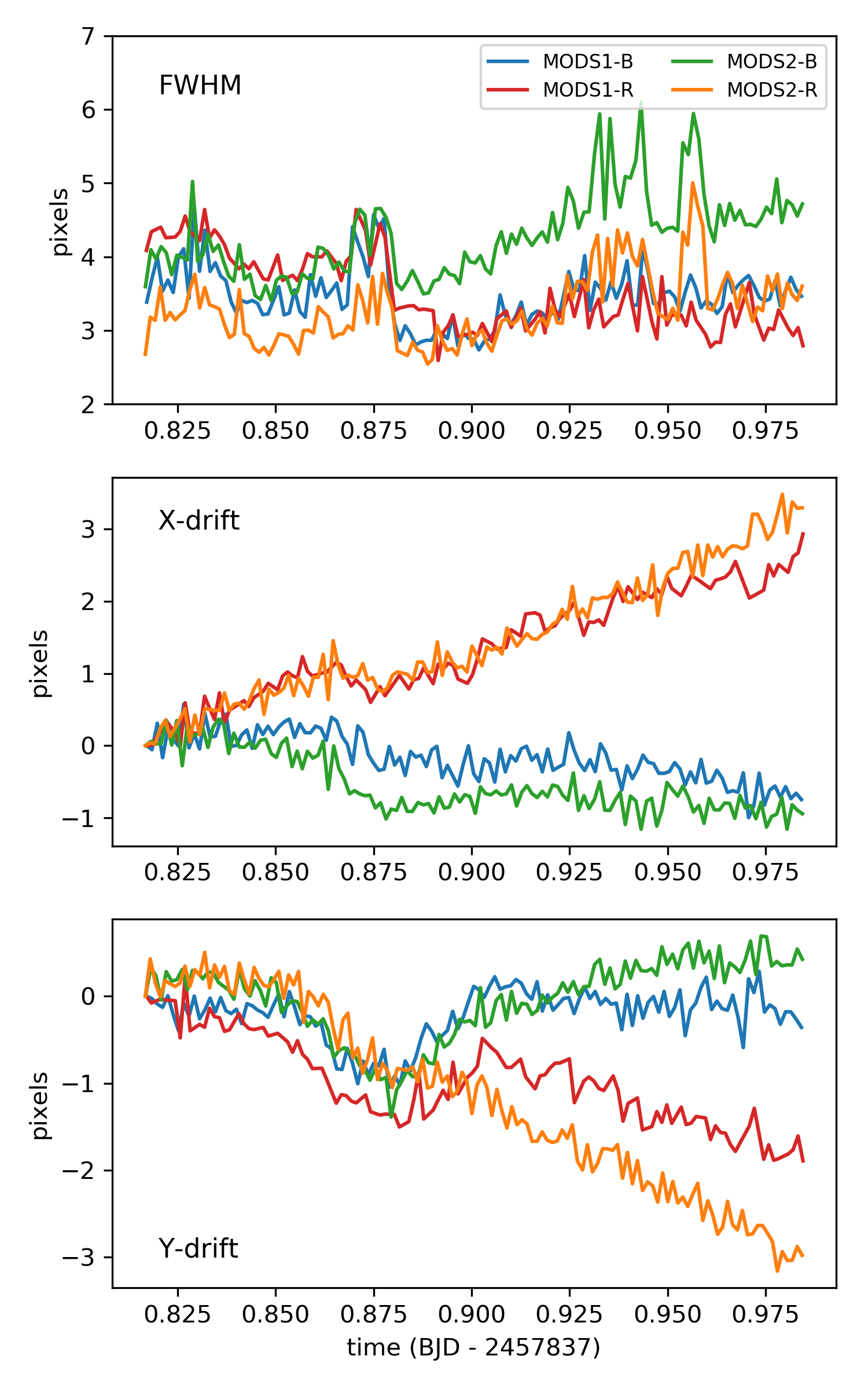}
      \caption{Time series of physical variables.
      \textit{Upper panel:} Measured FWHM of each channel.
      \textit{Middle panel:} X-direction drift (along the spectral dispersion direction).
      \textit{Bottom panel:} Y-direction drift (along the slit direction).
      }
         \label{FWHM}
   \end{figure}

We extracted the spectrum using different apertures that were centered at the spectral centroid. The centroid was calculated by applying a Gaussian fit along the spatial direction (i.e., the direction of the slit). 
We tested different aperture sizes ranging from 2 pixels to 40 pixels and calculated the root mean square (rms) of the residuals between the observed light curves and the best-fit models (the fit method is presented in Section 3). In this way, we identified the optimal aperture sizes that produce the lowest rms values. The optimal aperture sizes for the four channels (MODS1-B, MODS1-R, MODS2-B, and MODS2-R) are 14 pixels, 24 pixels, 20 pixels, and 14 pixels, respectively.
The sky background values were calculated using the median of two 10-pixel-wide regions located above and below the spectral aperture. The background region was chosen to be as far away from the object as possible to avoid contamination from the object flux \citep{Mallonn2016}. The objects drifted in both the spectral dispersion direction (X direction) and the slit direction (Y direction), probably because of the telescope pointing drift (Fig.~\ref{FWHM}). These drifts were measured and corrected for.

\begin{figure}
   \centering
   \includegraphics[width=0.45\textwidth]{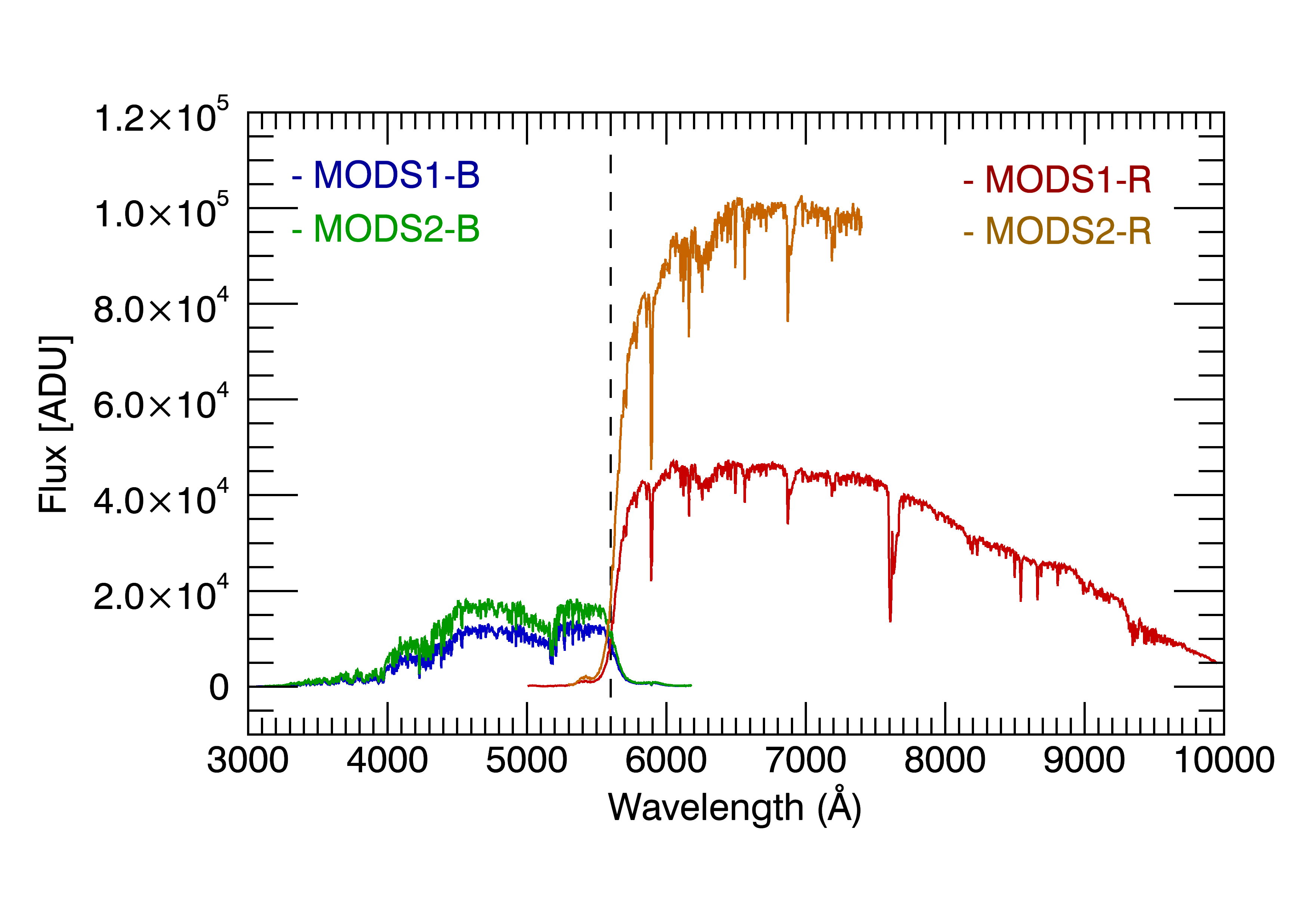}
      \includegraphics[width=0.45\textwidth]{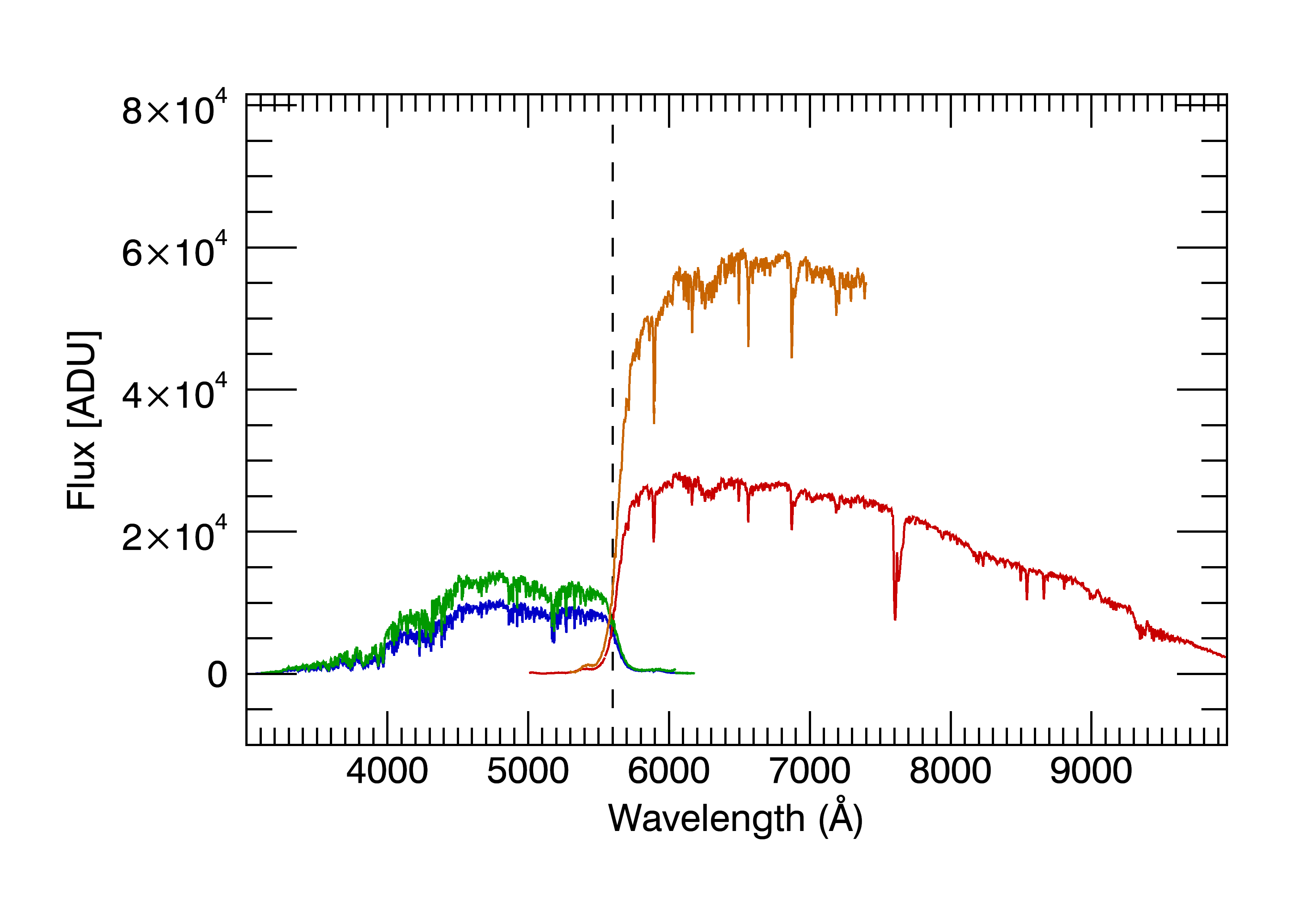}
      \caption{Example of the extracted spectra of HAT-P-12 (upper panel) and the comparison star (lower panel). These spectra were obtained simultaneously from the red and blue channels of the MODS1 and MODS2 spectrographs.}
         \label{raw-spectrum}
   \end{figure}

Fig.~\ref{raw-spectrum} shows an example of the final extracted spectra of the target star and the comparison star. 
The spectral flux of MODS2 is higher than the flux of MODS1 because the instrumental efficiency of MODS2 is higher than MODS1 by a factor of $\sim$ 2.5 at 7000 $\mathrm{\AA}$ and $\sim$ 1.4 at 5000 $\mathrm{\AA}$. The flux of the red channel is also higher than the blue channel because of their different efficiencies and because the telluric extinction at the blue wavelengths is stronger.
The MODS2-R spectrum at wavelengths larger than 7100 $\AA$ is affected by bad pixel columns, thus we only used the spectra with wavelengths below 7100 $\AA$ for the MODS2-R channel. The light curves were then obtained by integrating the flux values within the given wavelength ranges.

%

%

\section{Transit light-curve analysis}
\subsection{Fitting method}
We modeled the raw flux of the target star in magnitude space in order to convert the inherently multiplicative impact of systematics on the relative flux into an additive problem. We considered
here a model in which the time-varying target stars' magnitude can be explained by (1) a 
transit model \citep[here modeled using \texttt{batman};][]{Kreidberg2015}, (2) the 
(weighted) time-varying magnitude of the comparison star, and (3) a model that accounts for any 
systematic variation that is not captured by the comparison star or the transit model, such as changes 
in magnitude in the target star due to FWHM changes, centroid shifts, etc. 
These variables have correlated with the observed light curves.
We followed an approach that is very similar to that of \cite{gibson:2014}, in which we considered 
that this latter systematic model can be captured by a Gaussian process (GP). In summary,
our model for the magnitude of the target star is given by
\begin{equation}
m(t) = c_0 + Am^\textnormal{R}(t) -2.51\log_{10} T(t) + \epsilon,    
\end{equation}
where $m\textnormal(t)$ is the (mean-subtracted) magnitude of the target star, $m^\textnormal{R}(t)$ is the 
(mean-subtracted) magnitude of the comparison star, $c_0$ is a magnitude offset, $A$ is a weight for the comparison 
star, $T(t)$ is a transit model, and $\epsilon$ is a stochastic component, here modeled as a GP, that is, $\epsilon \sim N(\mathbf{0},\mathbf{\Sigma})$, with a covariance matrix defined by 
$\mathbf{\Sigma}_{i,j} = k(\mathbf{x}_i,\mathbf{x}_j) + \sigma^2_w\delta_{i,j}$. Here, $\sigma_w^2$ is simply a jitter term and $\delta_{i,j}$ is a Kroenecker delta, while $k(\mathbf{x}_i,\mathbf{x}_j)$ is modeled 
using a multidimensional squared-exponential kernel of the form
\begin{equation}
k(\mathbf{x}_i,\mathbf{x}_j) = \sigma^2_\textnormal{GP}\exp \left(-\sum^D_{d=1} \alpha_d(x_{d,i}-x_{d,j})^2\right),    
\end{equation}
where $\sigma_\textnormal{GP}$ is the amplitude of the GP and the $\alpha_d$ are 
the inverse (squared) length-scales of each of the components of the GP. The 
$\mathbf{x}_i$ vectors here have components $x_{d,i}$, where each $i$ denotes a time-stamp 
and where each $d$ corresponds to a different external variable. In our case, we considered time, 
FWHM, and X and Y centroid positions as possible external variables in our GP framework. 
These variables are found to have correlations with the observed light curves.
The variables were standardized (i.e., mean-subtracted and divided by their standard deviations) before we fed them into the GP.

Based on this model, the fitting parameters can thus be divided between the 
parameters of the transit model and those that were fit to account for the atmospheric and 
instrumental systematics. For the transit model, the fit parameters are the mid-transit time ($T_\mathrm{mid}$), impact parameter ($b$), scaled semimajor axis ($a/R_*$), planet-to-star radius ratio ($R_\mathrm{p}/R_*$), and the corresponding limb-darkening coefficients. The planetary orbital inclination ($i$) is then derived using the equation $ b = (a/R_*)~\mathrm{cos}~i $.
We used a quadratic limb-darkening law for the white-light curves and a linear law for the 
spectroscopic light curves (see below). For the white-light fits, we used the uninformative sampling scheme outlined in \cite{kipping:2013} and thus fit for the parameters $q_1$ and $q_2$ with uniform priors between 0 and 1, which were then 
converted into the limb-darkening coefficients $u_1$ and $u_2$ using the transformations presented in that work. The priors used in this work for these parameters are presented in Table \ref{priors-h12}. For the priors 
for the instrumental systematics, we set a large uniform prior between -2 and 2 for $c_0$, a large uniform prior between -10 and 10 for the weight of the comparison star ($A$) wide log-uniform priors for $\sigma_w$ and $\sigma_\textnormal{GP}$ between 0.01 and 100 mmag, and exponential 
priors with unitary scale for the $\alpha_d$.  We used \texttt{george} \citep{george} to 
evaluate the log-likelihood of our GP-based regression and perform the posterior sampling using 
 importance nested sampling via the MultiNest algorithm \citep{MultiNest}. In particular, we used 
the PyMultiNest \citep{PyMultiNest} package to perform the posterior sampling.

\begin{table}
\caption{Priors used for our light-curve fitting. $N(\mu,\sigma^2)$ stands for a normal distribution 
with mean $\mu$ and variance $\sigma^2$. $TN(\mu,\sigma^2)$ stands for a truncated normal with the same location and scale parameters. Here the truncated normal for $a/R_*$ is truncated between 1 
and 100, and between 0 and 1 for $R_p/R_*$ and $b$. Reference 1 is \cite{Alexoudi2018}.}             
\label{priors-h12}      
\centering                          
\begin{tabular}{l c c}        
\hline\hline                 
Parameter       &        Distribution & Reference \\     
        \hline                       
            &   \\
Period (days) & $N(3.21305766,0.00000013^2)$ & 1\\
$a/R_*$  & $TN(11.68,1.0^2)$ & 1\\
$R_p/R_*$  & $TN(0.138,0.1^2)$ & 1\\
$b$  & $TN(0.28,0.1^2)$ & 1\\
$q_1$  & $U(0,1)$ & ---\\
$q_2$  & $U(0,1)$ & ---\\
$T_\mathrm{mid}-2457837$  & $N(0.89,0.1^2)$ & ---\\
\hline                                   
\end{tabular}
\end{table}

\subsection{White-light curves}
\begin{figure*}
   \centering
   \includegraphics[width=0.90\textwidth]{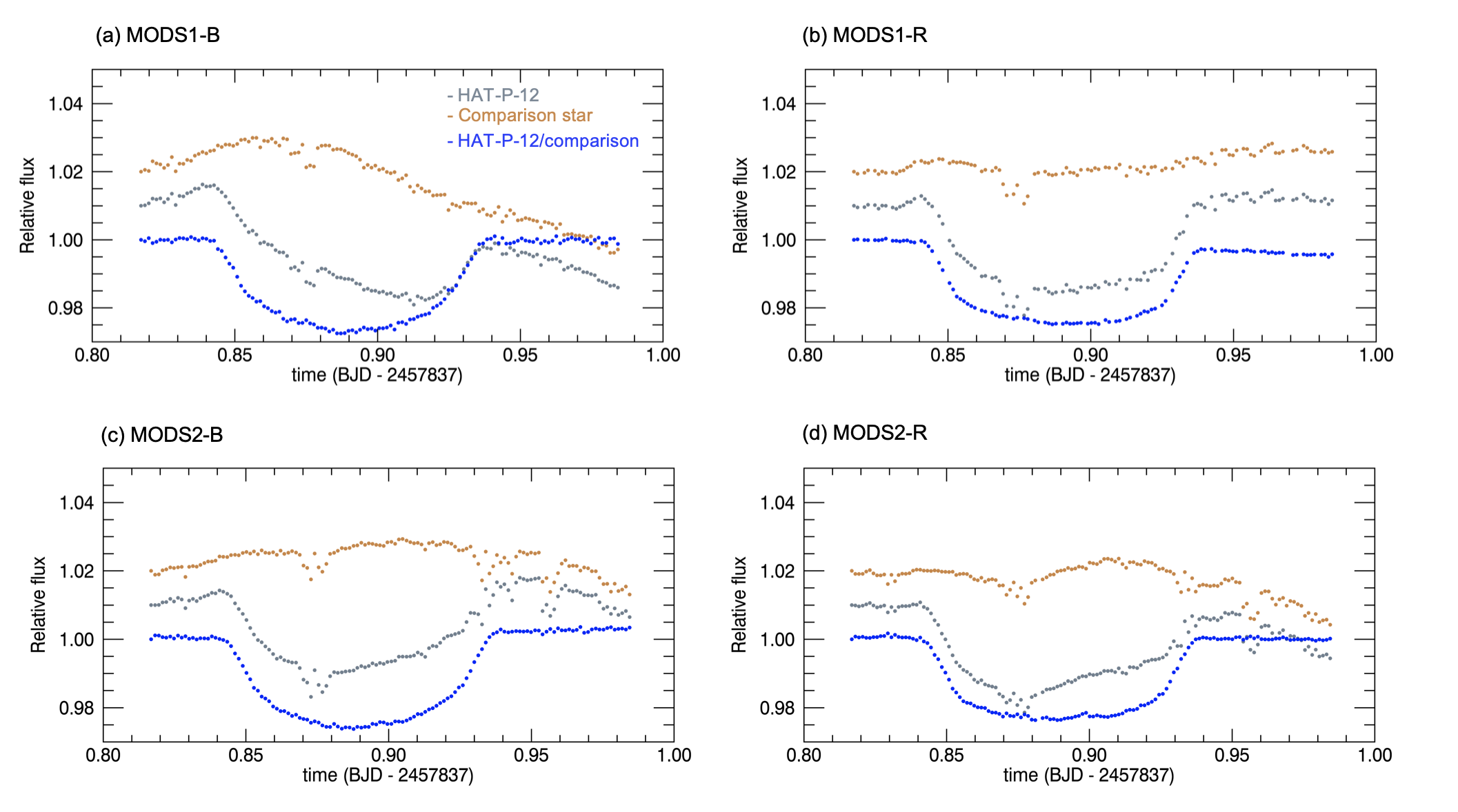}
      \caption{Raw white-light curves of the target star (gray points) and the comparison star (yellow points) for the four channels. The blue points are the ratio between the target star and the comparison star. The light curves are the normalized fluxes with constant offsets.}
         \label{Fig-T1C1}
   \end{figure*}

The transit light curves from the four channels were analyzed independently. We obtained the white-light curve of each channel by integrating the flux within a large wavelength range (see the last row in Table \ref{White-LC-paras}). Fig.~\ref{Fig-T1C1} shows the raw white-light curves of the target and comparison stars for each channel.

\begin{figure*}
   \centering
   \includegraphics[width=0.95\textwidth]{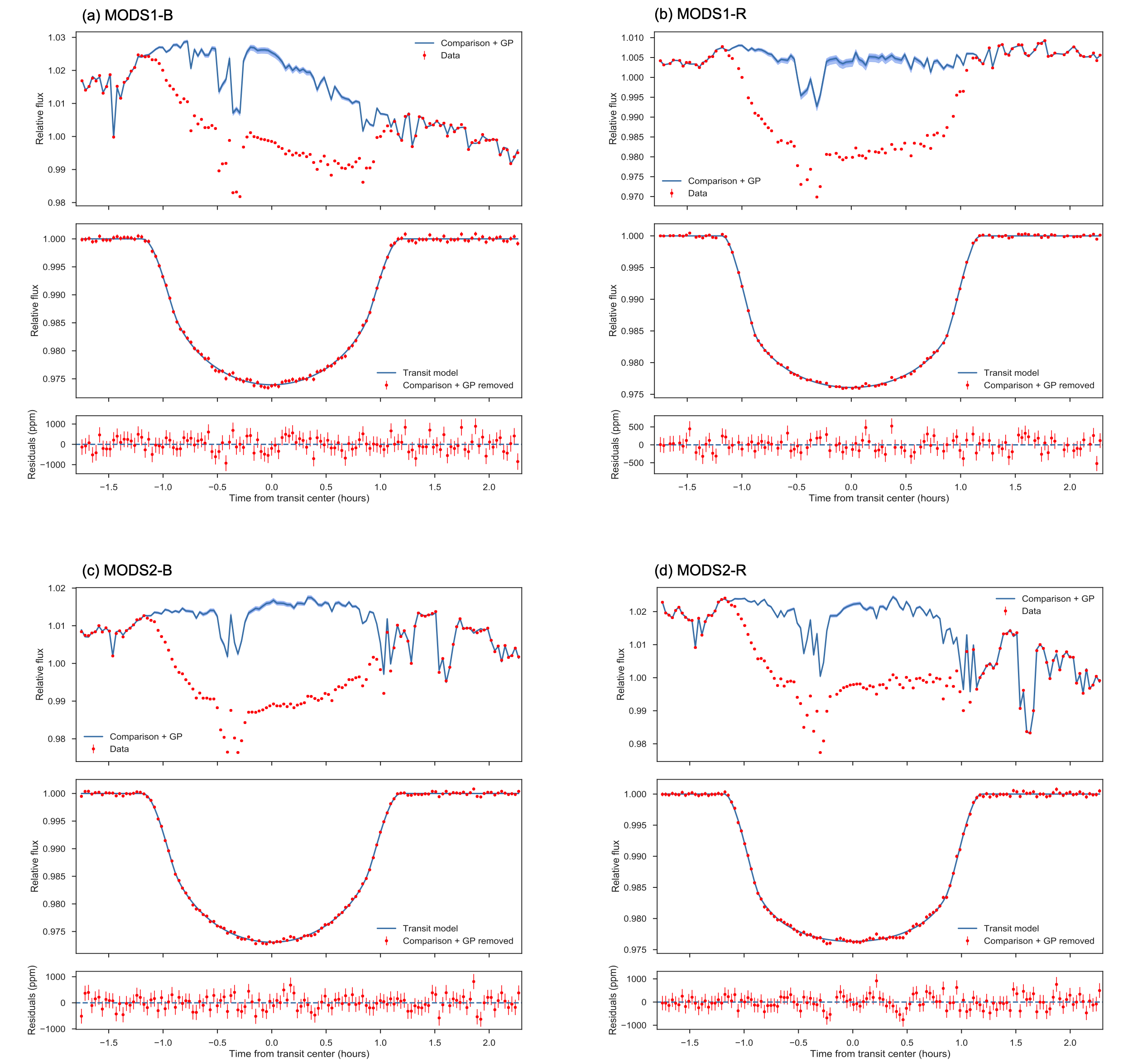}
      \caption{White-light curves for the four MODS channels.
      \textit{Upper:} Raw white-light curves of the target star together with the systematic models. Here the systematic model contains the light curve of the comparison star and the GP components. 
      \textit{Middle:} Light curve after the removal of the comparison star and GP components. The best-fit transit model is also presented.
      \textit{Bottom:} Residuals between the observation and model.}
         \label{WLC-figure}
   \end{figure*}

We fit the white-light curves using the fit method described above. A quadratic limb-darkening law was used (coefficients $u_1$ and $u_2$).
Fig.~\ref{WLC-figure} shows the white-light curves together with the best-fit models. 
Table \ref{White-LC-paras} presents the derived white-light curve parameters. The $T_\mathrm{mid}$, $i$, and  $a/R_*$ values from the four channels agree well within the uncertainties. We then averaged the values from the four channels to obtain the combined parameters. The combined $i$ and  $a/R_*$ values agree well with the results from A2018.

%

%

%
\begin{table*}
\caption{Measured parameters from the white-light curves of the four channels.}             
\label{White-LC-paras}      
\centering                          
\begin{tabular}{l c c c c c}        
\hline\hline                 
Parameter       &        MODS1-Blue      &      MODS1-Red & MODS2-Blue & MODS2-Red & combined \\     
        \hline                       
         $T_\mathrm{mid} [\mathrm{BJD} - 2457837] $)    & $0.88987_{-0.000113}^{+0.000106}$  &      $0.88975_{-0.000067}^{+0.000067}$  & $0.88975_{-0.000083}^{+0.000082}$ & $0.88974_{-0.000075}^{+0.000074}$ & $0.88977_{-0.000084}^{+0.000106}$ \\                 
                $a/R_*$         & $11.51_{-0.16}^{+0.16}$                       &       $11.67_{-0.10}^{+0.11}$  & $11.67_{-0.12}^{+0.11}$ & $11.55\pm0.12$ &   $11.61_{-0.15}^{+0.13}$ \\
                $i$ [degree]            & $88.73_{-0.28}^{+0.32}$                       &       $88.81_{-0.20}^{+0.23}$   & $89.02_{-0.26}^{+0.30}$ & $88.65_{-0.21}^{+0.26}$ & $88.80_{-0.25}^{+0.31}$ \\
        $R_\mathrm{p}/R_*$      & $0.1375_{-0.0016}^{+0.0017}$  &       $0.1380_{-0.0013}^{+0.0012}$  & $0.1391_{-0.0013}^{+0.0014}$ & $0.1382_{-0.0013}^{+0.0011}$ & $0.1382_{-0.0015}^{+0.0014}$ \\
        $u_1$                           & $0.87_{-0.12}^{+0.09}$  &     $0.47_{-0.07}^{+0.08}$  & $0.83_{-0.12}^{+0.11}$ & $0.64_{-0.11}^{+0.13}$ \\
        $u_2$                           & $0.47_{-0.04}^{+0.05}$  &     $0.47_{-0.07}^{+0.08}$  & $0.48_{-0.04}^{+0.06}$ & $0.29_{-0.05}^{+0.06}$ \\        
        rms [ppm]               & 356  & 190  & 283 & 259 \\
        $N$ (number of exposures) & 125  &      107  & 127 & 132 \\
        wavelength range        & 4000 - 5600 $\AA$  &  5600 - 8000 $\AA$  & 4000 - 5600 $\AA$ & 5600 - 6700 $\AA$ \\
\hline                                   
\end{tabular}
\end{table*}

\subsection{Spectroscopic light curves}
In order to obtain the transmission spectrum, we calculated spectroscopic light curves using similar wavelength bin sizes as in S2016. We fit these light curves as described in Section 3.1. We fixed the $T_\mathrm{mid}$, $i$, and  $a/R_*$ parameters to the average values obtained from the white-light curves.  According to \cite{Espinoza2016}, a linear limb-darkening law is as good as other laws when the noise level is above $\sim$ 1000 ppm. Thus we used a linear law for the spectroscopic light-curve fitting (coefficient $u_1$).

\begin{figure*}
   \centering
   \includegraphics[width=0.99\textwidth]{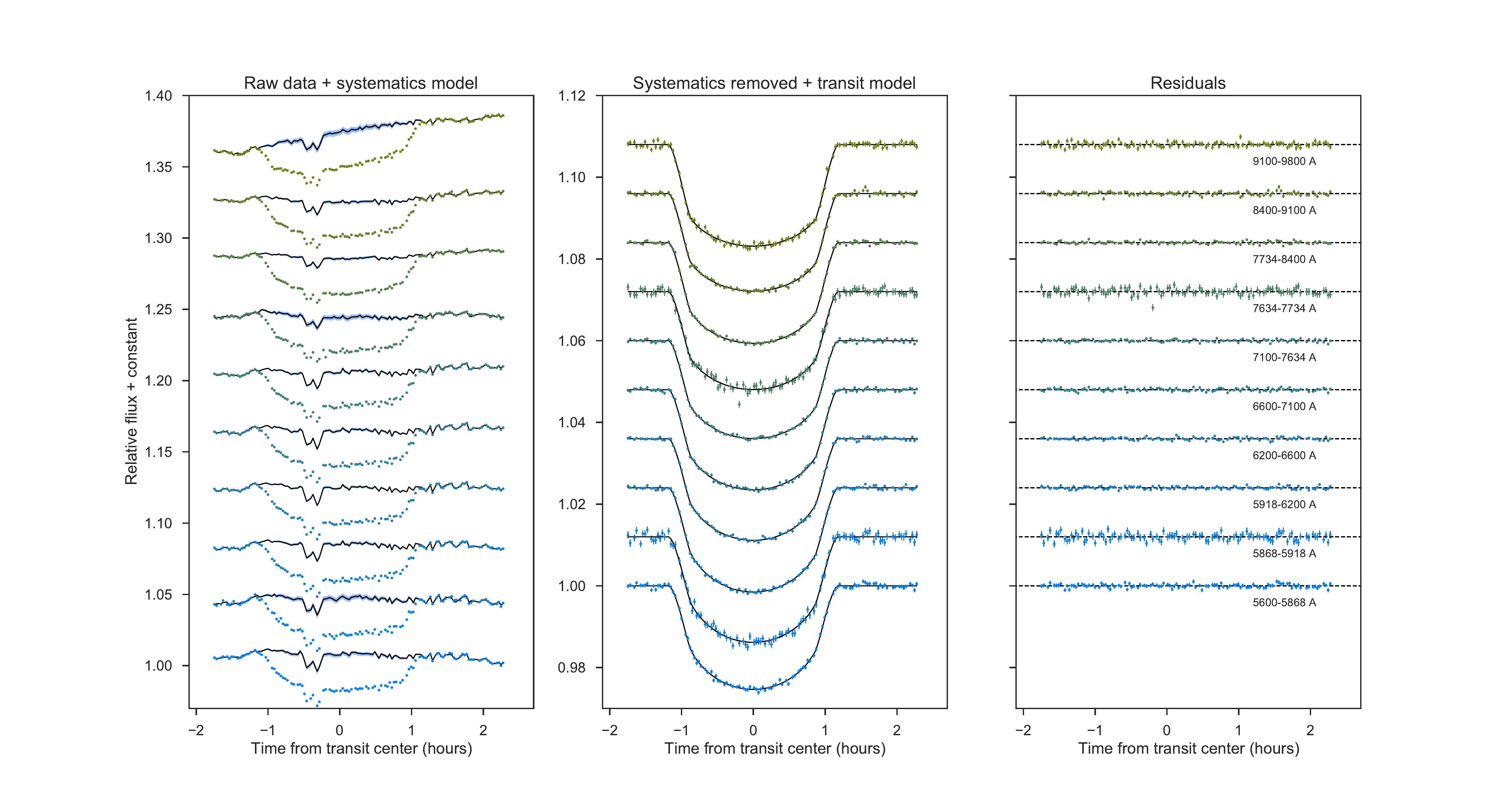}
   \includegraphics[width=0.99\textwidth]{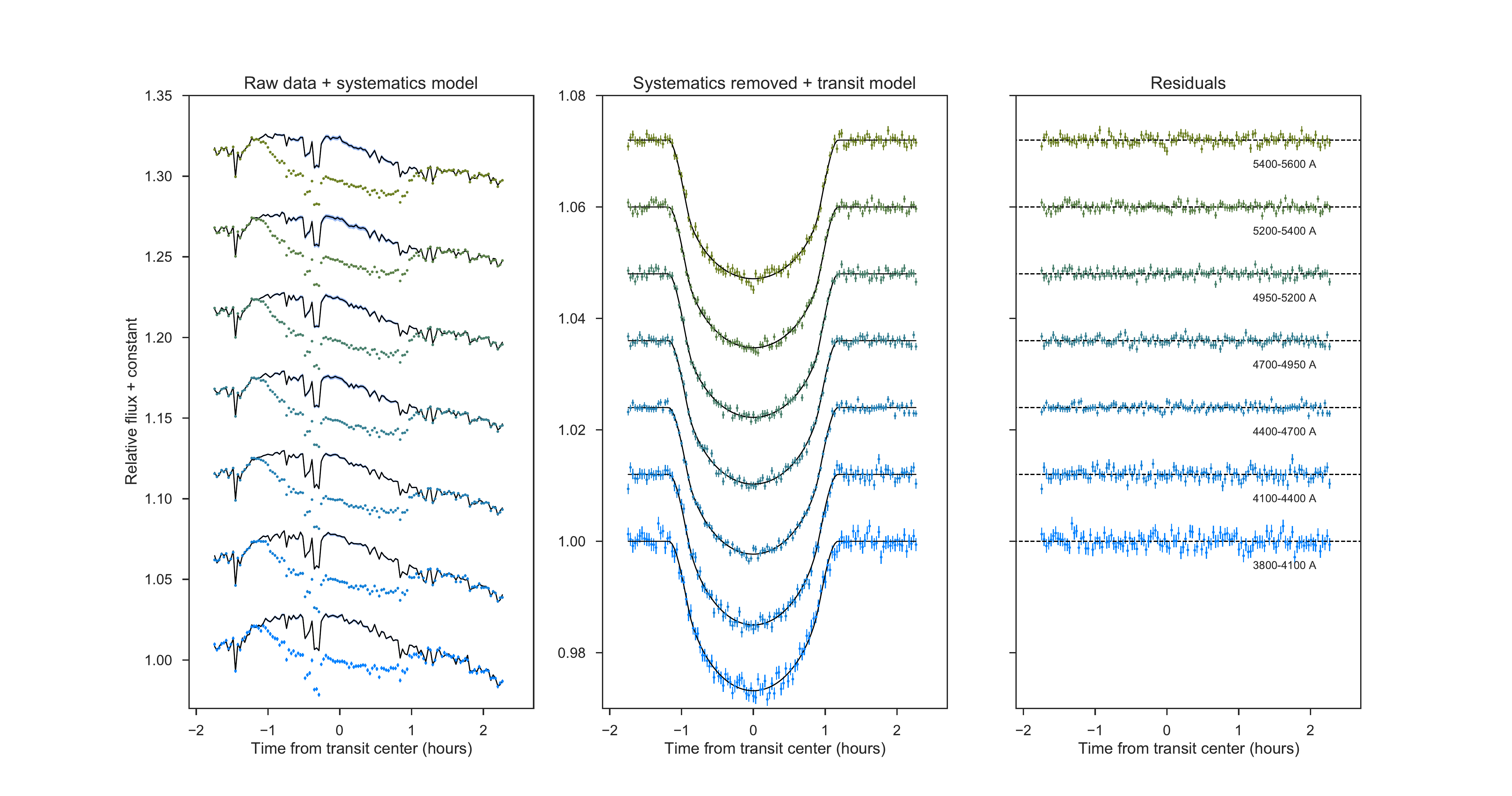}
      \caption{Wavelength-dependent light curves obtained using broad bin sizes from the MODS1 dataset.
\textit{Left:} Raw white-light curves of the target star together with the systematic models.
      \textit{Middle:} Light curves after the removal of the comparison star and GP components. The best-fit transit models are also plotted.
      \textit{Right:} Residual between the observation and model. The distance between two ticks on the vertical scale is 0.02.
       }
         \label{Fig-broad-LC-M1}
   \end{figure*}

\begin{figure*}
   \centering
         \includegraphics[width=0.99\textwidth]{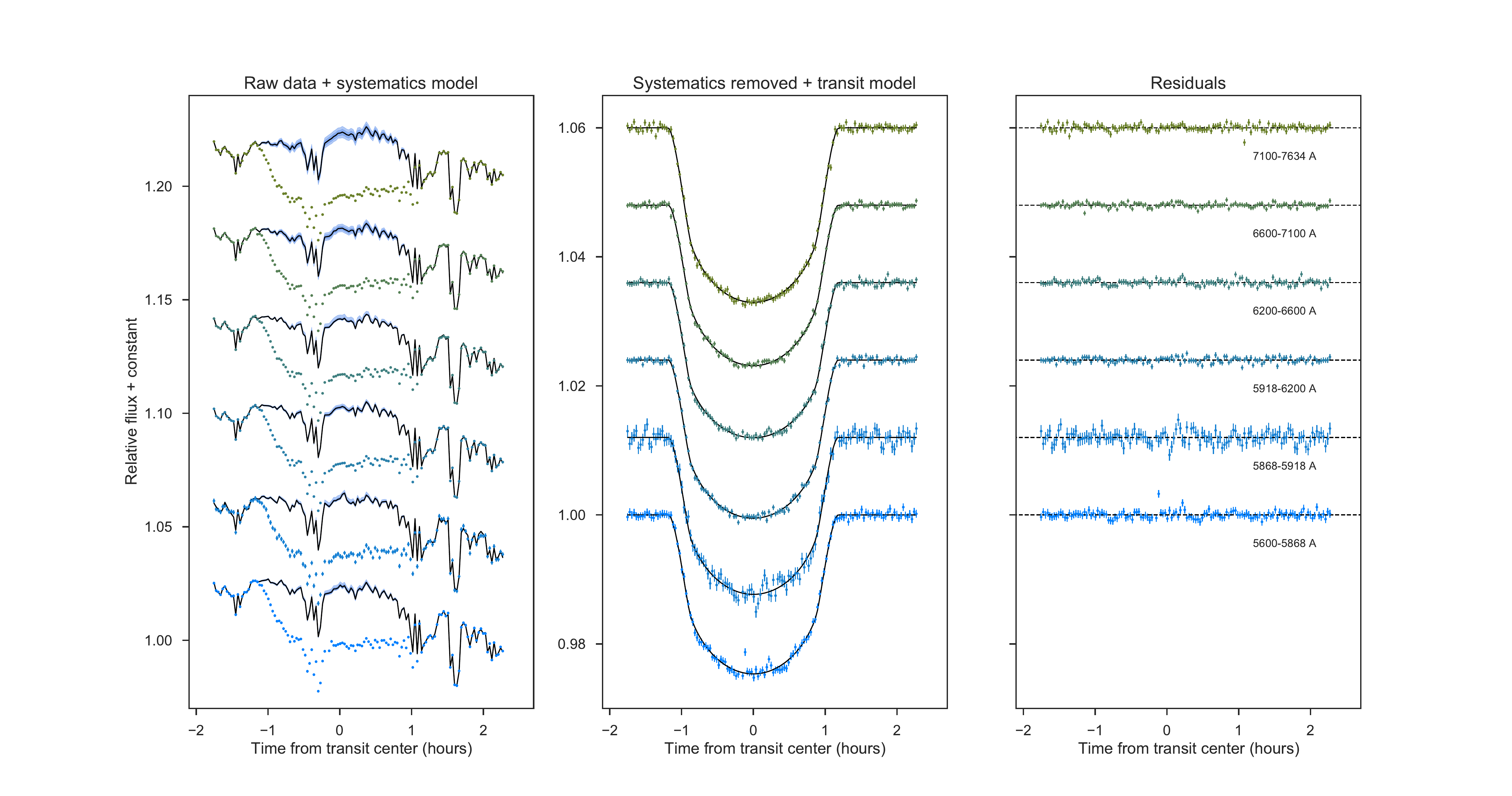}         
         \includegraphics[width=0.99\textwidth]{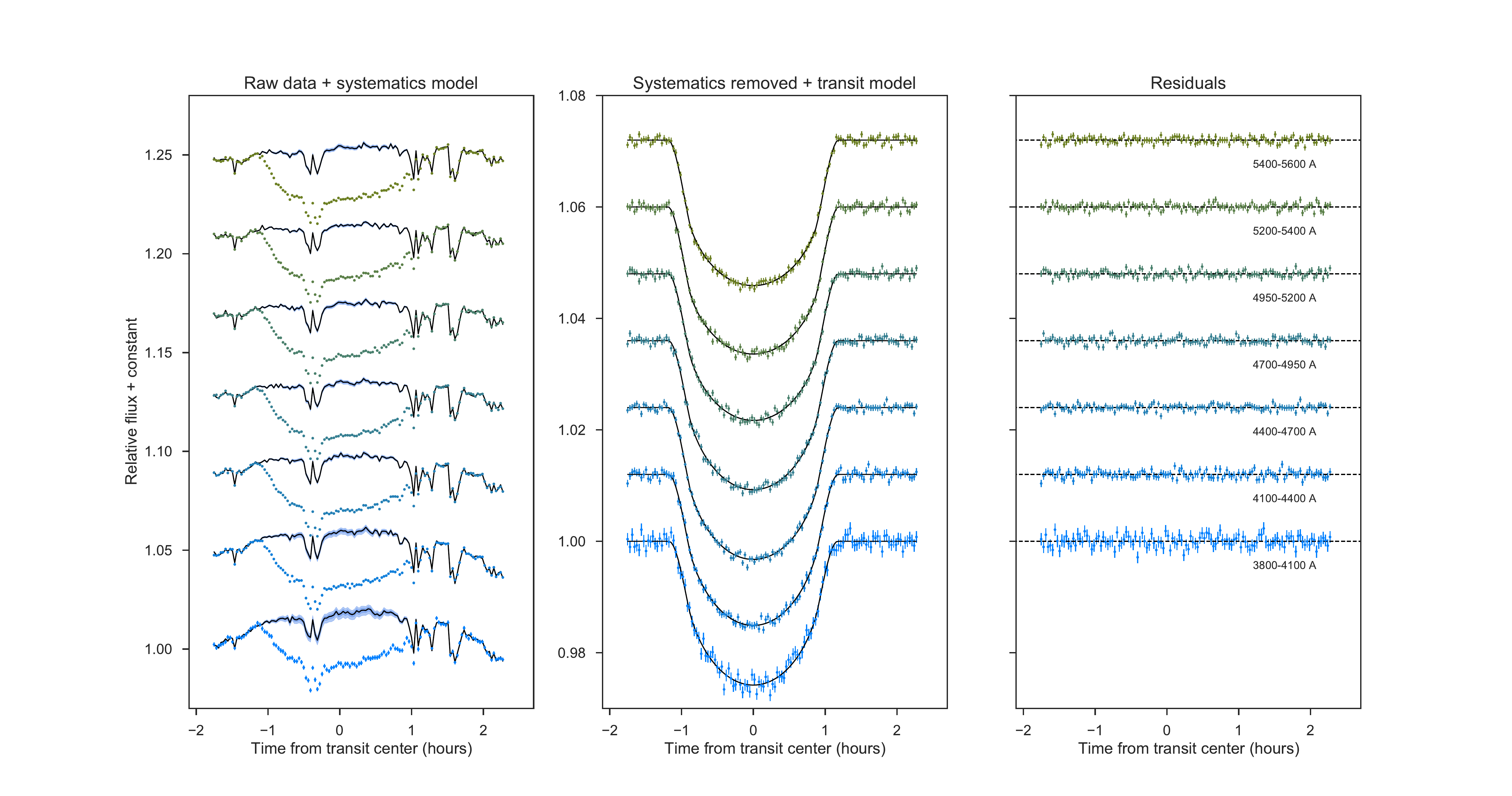}
      \caption{Same as Fig.~\ref{Fig-broad-LC-M1}, but for broad bins from MODS2 dataset.  
       }
         \label{Fig-broad-LC-M2}
   \end{figure*}

All the light curves together with the best-fit models are plotted in Figs.~\ref{Fig-broad-LC-M1} and \ref{Fig-broad-LC-M2}. The wavelength range of each band is shown next to the light curve. Table \ref{Tab-broad} summarizes the results of the fitted parameters, $R_\mathrm{p}/R_*$ and $u_1$. We analyzed the data in a wavelength range of $\sim$ 3800--9800 $\mathrm{\AA}$. However, at the blue and red ends of the overall wavelength coverage, the detector efficiency drops significantly and the telluric absorption is also prominent. Therefore we did not use the first data point in the blue channel and the last data point in the MODS1-R channel in the following analysis of the transmission spectrum.

In addition to these broad wavelength bins, we also calculated light curves using narrow bin sizes to search for potential Na and K absorptions.
We used a 50 $\AA$ bin centered at 5893 $\AA$ (the middle wavelength of the Na D doublet) and a 100 $\AA$ bin centered at 7684 $\AA$ (the middle wavelength of the two potassium lines).
The MODS1-R channel covers the wavelengths of both Na and K, while  MODS2-R only covers the wavelength of Na. We calculated nine narrow-band bins around the Na feature and five bins around the K feature. The fit results are shown in Table \ref{Tab-narrow}.

%

%

%
\begin{table*}[]
\centering
\caption{Fitting results of the broadband light curves.}
\label{Tab-broad}
\begin{tabular}{ccccccc}
\hline\hline                 
Band center &    Band size              & MODS1 &      & MODS2   &     &                   \\
($\AA$)   & ($\AA$)   & $R_\mathrm{p}/R_*$ & $u_1$  & $R_\mathrm{p}/R_*$     & $u_1$                    \\
        \hline                       
\textit{blue channel} \\
3950     &      300      &      0.1371 $^{+0.001}_{-0.001}$      &      0.969 $^{+0.018}_{-0.022}$     &      0.1345 $^{+0.0027}_{-0.003}$     &      0.972 $^{+0.02}_{-0.035}$      \\
4250     &      300      &      0.1374 $^{+0.0008}_{-0.0009}$    &      0.976 $^{+0.013}_{-0.018}$     &      0.1387 $^{+0.0015}_{-0.0018}$    &      0.939 $^{+0.024}_{-0.027}$     \\
4550     &      300      &      0.1375 $^{+0.0008}_{-0.0008}$    &      0.913 $^{+0.018}_{-0.02}$      &      0.1404 $^{+0.001}_{-0.001}$      &      0.897 $^{+0.018}_{-0.018}$     \\
4825     &      250      &      0.1368 $^{+0.0012}_{-0.0013}$    &      0.887 $^{+0.021}_{-0.021}$     &      0.1388 $^{+0.0011}_{-0.0012}$    &      0.908 $^{+0.019}_{-0.019}$     \\
5075     &      250      &      0.1387 $^{+0.0012}_{-0.0011}$    &      0.824 $^{+0.02}_{-0.021}$      &      0.1395 $^{+0.001}_{-0.0011}$     &      0.847 $^{+0.019}_{-0.02}$      \\
5300     &      200      &      0.1366 $^{+0.0014}_{-0.0016}$    &      0.851 $^{+0.022}_{-0.023}$     &      0.1412 $^{+0.0008}_{-0.0009}$    &      0.799 $^{+0.017}_{-0.018}$     \\
5500     &      200      &      0.1369 $^{+0.0013}_{-0.0013}$    &      0.802 $^{+0.025}_{-0.025}$     &      0.1403 $^{+0.0011}_{-0.0011}$    &      0.801 $^{+0.02}_{-0.021}$      \\
        \hline                       
\textit{red channel} \\
5734     &      268      &      0.139 $^{+0.0013}_{-0.0017}$     &      0.772 $^{+0.018}_{-0.017}$     &      0.1384 $^{+0.001}_{-0.0009}$     &      0.732 $^{+0.034}_{-0.03}$      \\
5893     &      50       &      0.1408 $^{+0.0023}_{-0.0022}$    &      0.759 $^{+0.032}_{-0.034}$     &      0.1377 $^{+0.001}_{-0.001}$      &      0.724 $^{+0.029}_{-0.024}$     \\
6059     &      282      &      0.1402 $^{+0.0012}_{-0.0011}$    &      0.749 $^{+0.014}_{-0.015}$     &      0.1385 $^{+0.0012}_{-0.001}$     &      0.712 $^{+0.022}_{-0.022}$     \\
6400     &      400      &      0.139 $^{+0.0008}_{-0.0009}$     &      0.73 $^{+0.013}_{-0.013}$     &      0.138 $^{+0.0009}_{-0.0006}$     &      0.685 $^{+0.025}_{-0.018}$     \\
6850     &      500      &      0.1394 $^{+0.0011}_{-0.0011}$    &      0.679 $^{+0.016}_{-0.016}$     &      0.139 $^{+0.0021}_{-0.002}$      &      0.732 $^{+0.024}_{-0.026}$     \\
7367     &      534      &      0.1382 $^{+0.0011}_{-0.0011}$    &      0.663 $^{+0.016}_{-0.016}$     &      0.1463 $^{+0.0025}_{-0.0026}$    &      0.687 $^{+0.031}_{-0.035}$     \\
7684     &      100      &      0.1388 $^{+0.0021}_{-0.0021}$    &      0.643 $^{+0.039}_{-0.039}$     &               &               \\
8067     &      666      &      0.1411 $^{+0.001}_{-0.001}$      &      0.632 $^{+0.014}_{-0.015}$     &               &               \\
8750     &      700      &      0.1398 $^{+0.001}_{-0.0012}$     &      0.593 $^{+0.018}_{-0.019}$     &               &               \\
9450     &      700      &      0.1437 $^{+0.0023}_{-0.0024}$    &      0.557 $^{+0.039}_{-0.047}$     &               &               \\
        \hline                       
\end{tabular}
\end{table*}

%
\begin{table*}[]
\centering
\caption{Fitting results of narrow bands adjacent to Na and K.}
\label{Tab-narrow}
\begin{tabular}{ccccccc}
\hline\hline                 
Band center &    Band size              & MODS1 &         &   MODS2    &       &                   \\
($\AA$)   & ($\AA$)   & $R_\mathrm{p}/R_*$ & $u_1$  & $R_\mathrm{p}/R_*$     & $u_1$                    \\
        \hline                       
Na &   &  &  &  &  &  \\
5693     &      50       &      0.1421 $^{+0.0024}_{-0.0025}$    &      0.749 $^{+0.032}_{-0.034}$     &      0.1395 $^{+0.0015}_{-0.0013}$    &      0.67 $^{+0.031}_{-0.033}$     \\
5743     &      50       &      0.1357 $^{+0.0025}_{-0.0028}$    &      0.851 $^{+0.032}_{-0.034}$     &      0.1381 $^{+0.0013}_{-0.0015}$    &      0.679 $^{+0.039}_{-0.037}$     \\
5793     &      50       &      0.1356 $^{+0.0033}_{-0.0037}$    &      0.789 $^{+0.049}_{-0.051}$     &      0.137 $^{+0.0015}_{-0.0016}$     &      0.766 $^{+0.038}_{-0.036}$     \\
5843     &      50       &      0.1437 $^{+0.0027}_{-0.0027}$    &      0.71 $^{+0.038}_{-0.039}$     &      0.1366 $^{+0.0016}_{-0.0015}$    &      0.779 $^{+0.039}_{-0.039}$     \\
5893     &      50       &      0.1408 $^{+0.0022}_{-0.0022}$    &      0.758 $^{+0.032}_{-0.034}$     &      0.1377 $^{+0.001}_{-0.001}$      &      0.721 $^{+0.029}_{-0.024}$     \\
5943     &      50       &      0.1369 $^{+0.0019}_{-0.0023}$    &      0.765 $^{+0.032}_{-0.032}$     &      0.1377 $^{+0.001}_{-0.0009}$     &      0.707 $^{+0.033}_{-0.025}$     \\
5993     &      50       &      0.1403 $^{+0.0022}_{-0.0022}$    &      0.758 $^{+0.034}_{-0.034}$     &      0.1382 $^{+0.0015}_{-0.0013}$    &      0.702 $^{+0.04}_{-0.034}$      \\
6043     &      50       &      0.145 $^{+0.0026}_{-0.0026}$     &      0.769 $^{+0.033}_{-0.035}$     &      0.1387 $^{+0.0014}_{-0.0011}$    &      0.692 $^{+0.032}_{-0.028}$     \\
6093     &      50       &      0.1411 $^{+0.0023}_{-0.0023}$    &      0.742 $^{+0.035}_{-0.035}$     &      0.1387 $^{+0.0013}_{-0.0011}$    &      0.693 $^{+0.035}_{-0.029}$     \\
        \hline                       
K &   &  &  &  &  &  \\
7484     &      100      &      0.1405 $^{+0.0016}_{-0.0017}$    &      0.659 $^{+0.025}_{-0.026}$     &               &               \\
7584     &      100      &      0.1402 $^{+0.0028}_{-0.0027}$    &      0.633 $^{+0.043}_{-0.046}$     &               &               \\
7684     &      100      &      0.1388 $^{+0.0021}_{-0.0022}$    &      0.645 $^{+0.041}_{-0.04}$      &               &               \\
7784     &      100      &      0.1438 $^{+0.0015}_{-0.0015}$    &      0.662 $^{+0.023}_{-0.024}$     &               &               \\
7884     &      100      &      0.1421 $^{+0.0018}_{-0.0017}$    &      0.677 $^{+0.03}_{-0.028}$      &               &               \\
        \hline                       
\end{tabular}
\end{table*}

\section{Result and discussion}
\subsection{LBT results}
\begin{figure*}
   \centering
   \includegraphics[width=0.8\textwidth, height=0.35\textwidth]{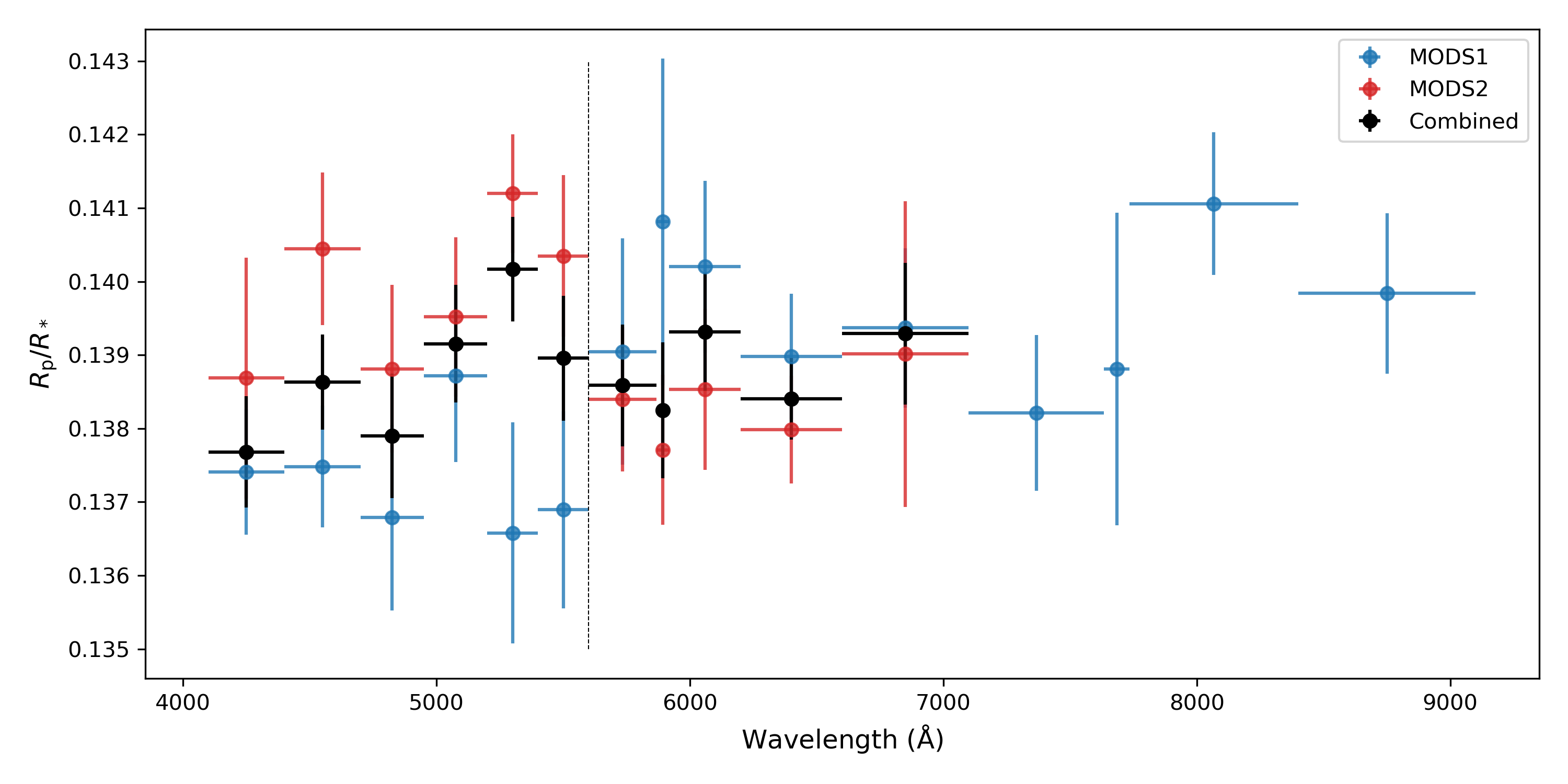}
      \caption{Transmission spectra of HAT-P-12b from the LBT/MODS observation. The blue and red points are the spectra from MODS1 and MODS2, respectively. The black points are the average spectrum. The vertical dashed line indicates the wavelength boundary between the blue and red channels.
      }
         \label{Fig-Tran-MODS1-2}
   \end{figure*}

\begin{figure}
   \centering
   \includegraphics[width=0.43\textwidth]{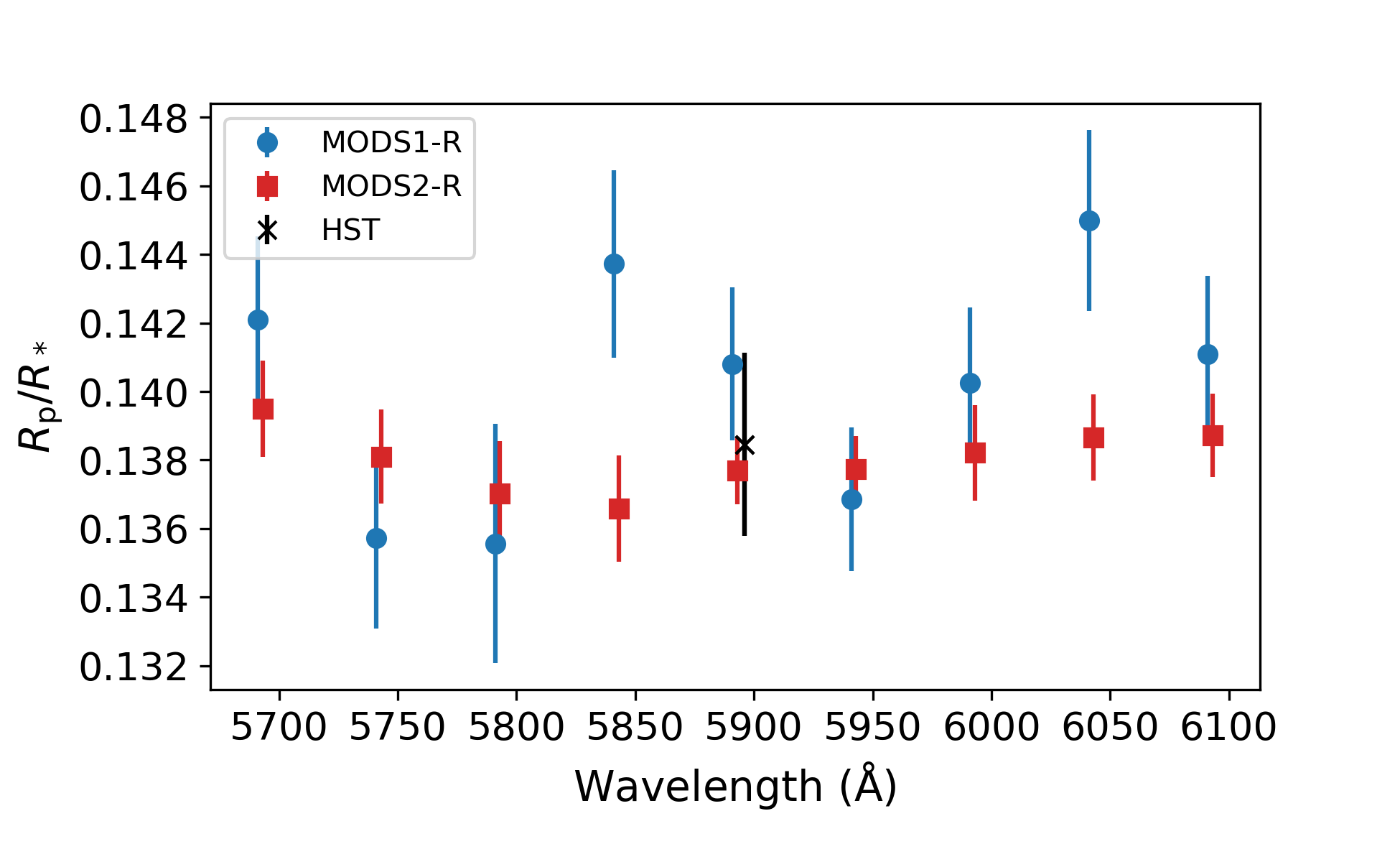}
      \includegraphics[width=0.43\textwidth]{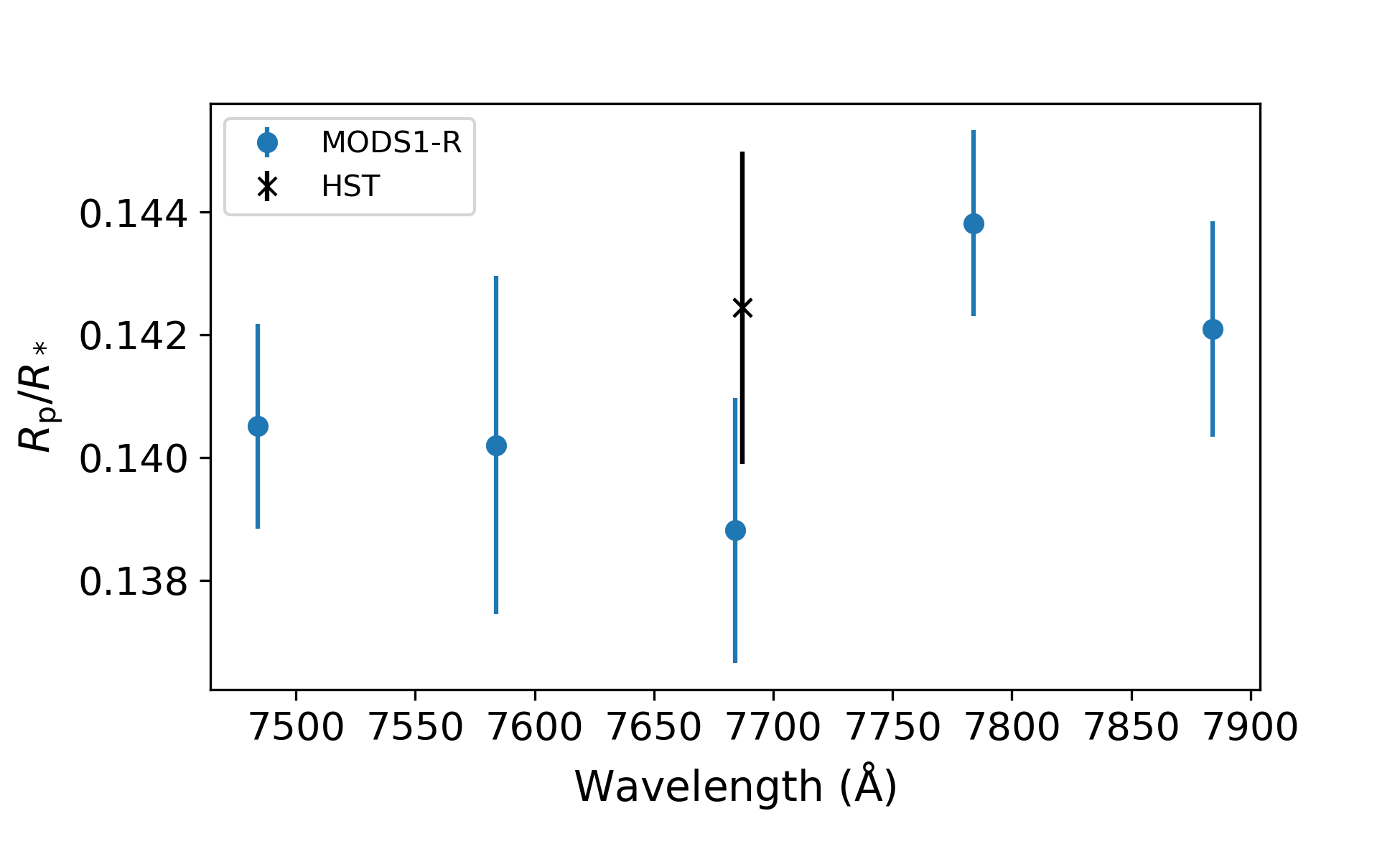}
      \caption{Transmission spectra around the Na feature (upper panel) and K feature (lower panel). The bin size is 50 $\AA$ for Na and 100 $\AA$ for K. The values of Na and K bins from the \textit{HST} observation (A2018) are also plotted as black points. The LBT results have similar or even smaller errors than the \textit{HST} data points. The spectra are in general flat. The wavelengths from different observations are slightly shifted for clarity.
            }
         \label{Fig-Tran-Na-K}
   \end{figure}

The broadband transmission spectra from the MODS1 and MODS2 observations are plotted in Fig.~\ref{Fig-Tran-MODS1-2}. The results from the two spectrographs are consistent with each other, and we subsequently averaged the MODS1 and MODS2 spectra with the inverse of the squared uncertainties as weights (black points in Fig.~\ref{Fig-Tran-MODS1-2}).
The narrow-band transmission spectra at Na and K wavelengths are presented in Fig.~\ref{Fig-Tran-Na-K}.

Both the broad- and the narrow-band transmission spectra from our LBT observations are flat, and we did not detect any Na/K absorption or a strong Rayleigh-scattering slope.
According to theoretical predictions \citep[e.g.,][]{Seager2000}, clear atmospheres of giant planets should have broad wings of Na and K and Rayleigh-scattering slopes toward the blue, and these features have been observed in some exoplanets \citep[e.g.,][]{Sing2008,Nikolov2018}.
Because no such broad Na/K wings are detected in our data, we are able to confirm with our LBT results that there are aerosols or clouds in the planetary atmosphere.

%

%

   \begin{figure*}
   \centering
   \includegraphics[width=0.8\textwidth, height=0.35\textwidth]{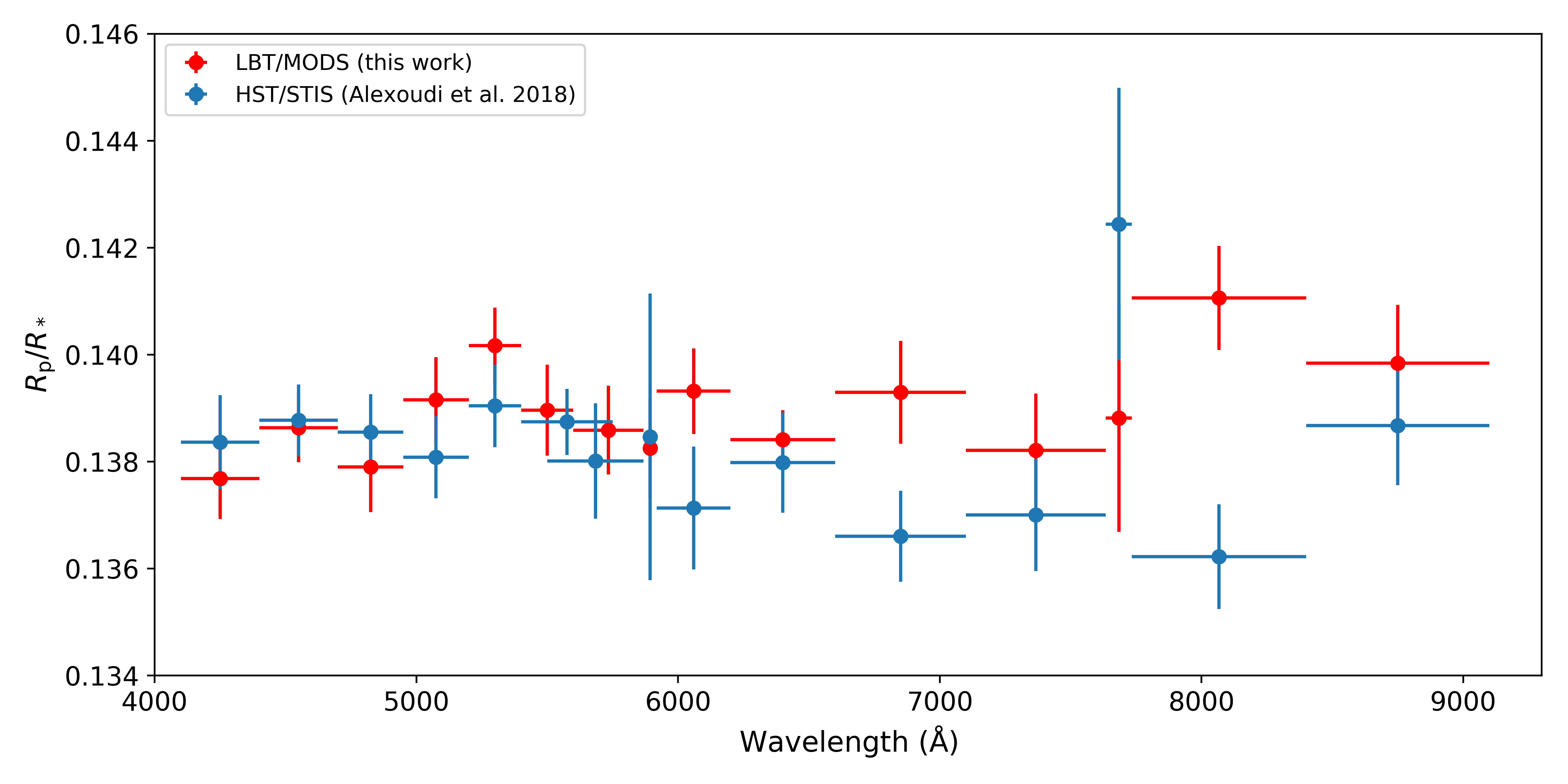}
      \caption{Comparison between the LBT/MODS result (red points) and the \textit{HST}/STIS result (blue points, A2018). 
      Our LBT result agrees well with the reanalyzed \textit{HST} result from A2018.
      }
         \label{Fig-LBT+HST}
   \end{figure*}

\subsection{Comparison with HST results in the visible}
S2016 observed three transits of HAT-P-12b with \textit{HST}/STIS in the visible wavelengths. They detected a strong Rayleigh-scattering slope and a sign of K absorption. 
A2018 reanalyzed the \textit{HST} data with updated planetary orbital parameters and obtained a relatively flat transmission spectrum. They concluded that the slope in S2016 is probably a result of using incorrect planetary orbital parameters.

The planetary orbital parameters obtained from our LBT observations ($a/R_* = 11.61_{-0.15}^{+0.13}$,
$i = 88.80_{-0.25}^{+0.31}$) are very similar to those in A2018 ($a/R_* = 11.68\pm{0.12}$,
$i = 88.83\pm{0.19}$). The LBT broadband transmission spectrum also agrees well with the A2018 result (Fig.~\ref{Fig-LBT+HST}). The average $R_\mathrm{p}/R_*$ value of the data points within the wavelength range of the MODS blue channel is 0.1387 for the LBT data and 0.1386 for the \textit{HST} data in A2018; while for the \textit{HST} data in S2016, the average $R_\mathrm{p}/R_*$ is 0.1415.
For the Na feature, both our results and the \textit{HST} results show no significant absorption. The \textit{HST} data point at the Na wavelength has a very similar value as the LBT data points (cf. the upper panel in Fig.~\ref{Fig-Tran-Na-K}).
Therefore we confirm the revised \textit{HST} result in A2018.

The potassium feature in the broadband transmission spectra (Fig.~\ref{Fig-LBT+HST}) shows a slight discrepancy between our LBT results and the \textit{HST} results (A2018). The LBT results suggest a nondetection of K, while the \textit{HST} result suggests a tentative detection of the K absorption.
However, when we compare the spectra obtained using narrow bin sizes around the K wavelength (lower panel in Fig.~\ref{Fig-Tran-Na-K}), we found that the \textit{HST} data point at the K wavelength is consistent with the LBT data points around the K wavelengths. These narrow-band data points have significant statistical fluctuations due to their relatively large errors. When we compare just the narrow K bin with the nearby broad spectral bins, as shown in Fig.~\ref{Fig-LBT+HST}, such a statistical fluctuation could result in an absorption-like feature. Therefore we recommend caution when narrow-band results from low spectral resolution observations are interpreted. Although the LBT results indicate a nondetection of potassium, we cannot confidentially rule out the existence of a weak K feature, considering the large errors of the light curves acquired with narrow spectral bins and the effect of telluric oxygen absorption adjacent to the K absorption lines.

\cite{Deibert2019} observed the transit of HAT-P-12b with the echelle high-dispersion spectrograph mounted on the Subaru telescope. These authors achieved a 3.2\,$\sigma$ detection of sodium absorption. However, they were not able to detect potassium at a statistically significant level, and they ruled out a K absorption feature down to an amplitude of 2\,$\%$ relative to the normalized flux.

Discrepancy in the potassium feature between \textit{HST} and ground-based observations also occurs for another exoplanet, \object{WASP-31b}.
\cite{Gibson2017} observed the transmission spectrum of WASP-31b with VLT/FORS2 and compared their result with the \textit{HST} result \citep{Sing2016}. The VLT/FORS2 spectrum does not show any significant detection of K, while the \textit{HST} result shows a K feature at high significance. 
\cite{Gibson2019} further observed the planet at high spectral resolution with the VLT/UVES spectrograph and confirmed the nondetection of K.
They attributed the discrepancy to the underestimated instrumental systematics of the \textit{HST}/STIS instrument around the K wavelengths. The potassium discrepancies between ground-based observations and \textit{HST} observations of HAT-P-12b and WASP-31b demonstrate the importance of repeating observations with different instruments in the study of exoplanet atmospheres.

\subsection{Effect of stellar activity}
The stellar disk is normally not homogeneous because of magnetically active regions such as spots and faculae. Such a photospheric heterogeneity may imprint stellar spectra features in the obtained transmission spectrum through the transit light source (TLS) effect.  For example, \citet{Rackham2017} obtained the visual transmission spectrum of GJ~1214b and showed that it receives a contribution from unocculted stellar faculae. Below we discuss the effect of stellar activity on the transmission spectrum.

\subsubsection{Photometric monitoring program }
We monitored HAT-P-12 in two colors (B and V bands) with the STELLA Robotic Observatory and its wide-field imager WiFSIP \citep{Strassmeier2004} from January to July 2017 to investigate photometric variations caused by stellar activity. Details of the monitoring program are described in \cite{Mallonn2015}. 

\begin{figure}
   \centering
   \includegraphics[width=0.49\textwidth]{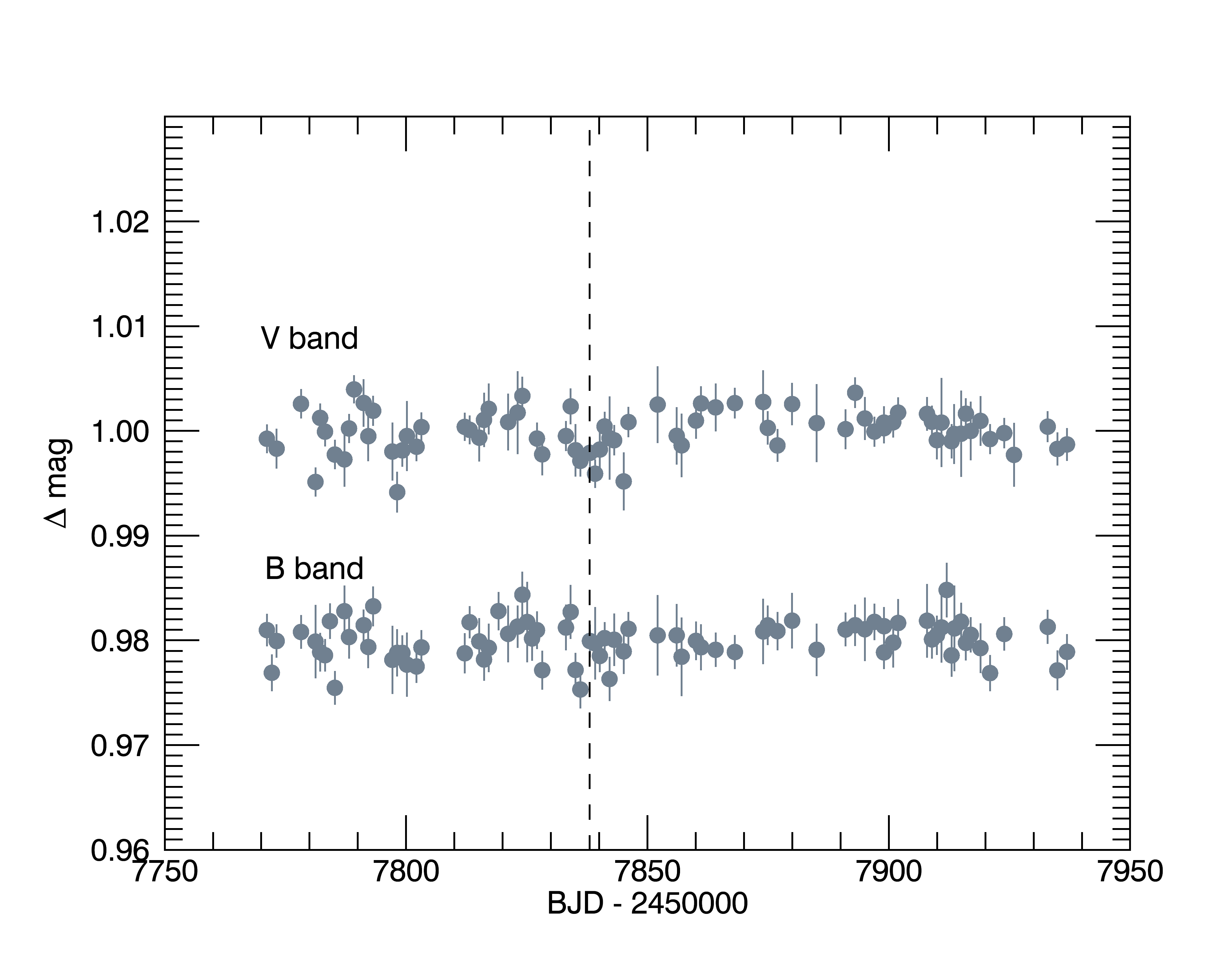}
      \caption{Photometry monitoring data points of HAT-P-12 in 2017. The date when the LBT observation was performed is indicated as a vertical dashed line.
      }
         \label{Photometry}
   \end{figure}

The light curves are plotted in Fig.~\ref{Photometry}, and the LBT observation date are indicated as a vertical line in the figure. The light curves are essentially flat, with a standard deviation of 1.8\,mmag for B band and 2.0\,mmag for V band. The Lomb-Scargle periodogram shows no significant period, which agrees with the result in \cite{Mancini2018}. Thus, the photometric monitoring suggests that the star was not very active during the monitoring period.

\subsubsection{Stellar contamination model}
Using the 2017 light curves, we estimated the spot-covering fraction and the associated TLS signals (i.e., stellar spectral contamination) following the approach of \citet{Rackham2018}.
In short, the approach consists of (1) using an ensemble of model stellar photospheres with active regions added at random locations to estimate the active region-covering fraction corresponding to an observed variability amplitude, and (2) calculating the maximum TLS signals they would produce if no active regions were occulted by the transiting planet (i.e., a worst-case scenario for the TLS effect).

In the photospheric modeling, we used full-disk stellar spectra interpolated from the PHOENIX model grid \citep{PHOENIX} to approximate the emergent spectra of the quiescent photosphere, spots, and faculae.
We used model spectra with $\log g=4.61$ and $\mathrm{[Fe/H]}=-0.29$, following \citet{Hartman2009}.
We set the temperatures of the quiescent photosphere, spots, and faculae to 4650\,K, 3560\,K, and 4750\,K, respectively, using the stellar effective temperature \citep{Hartman2009} as the quiescent photosphere temperature and following the scaling relations of \citet{Rackham2019} for the spot and facula temperatures.
Following this work, we adopted spot radii of $R_\mathrm{spot}=2\,\degr$ and a 10:1 facula-to-spot areal ratio.

With this approach, we estimated the spot coverages consistent with the $B$-band and $V$-band light curves for models with only spots and models with both spots and faculae.
Because the light curves show no significant periodicity, we used the standard deviations of the photometry data points as the light-curve variation semiamplitudes.
For the spots-only models, we find consistent spot coverages for both bands and conservatively report here the higher value $0.5^{+0.6}_{-0.3}\%$.
If such a spot coverage were present in the unocculted stellar disk during the transit, we estimated that it could increase the observed $R_\mathrm{p}/R_*$ values by $0.2^{+0.2}_{-0.1}\%$ or ${\sim}0.0002\,R_\mathrm{p}/R_*$ on average over the wavelength range of the LBT observations.
For the models with both spots and faculae, the net effect of the estimated spot and facula coverages would be to decrease the observed $R_\mathrm{p}/R_*$ values by $0.3^{+0.1}_{-0.1}\%$ or ${\sim}0.0004\,R_\mathrm{p}/R_*$ on average.
In either case, such a change is well below the uncertainties of our obtained transmission spectrum.
We therefore conclude that the contribution of TLS signals to the observed transmission spectrum can be neglected.

\subsection{Planetary atmosphere model}
\begin{figure*}
   \centering
   \includegraphics[width=0.8\textwidth, height=0.4\textwidth]{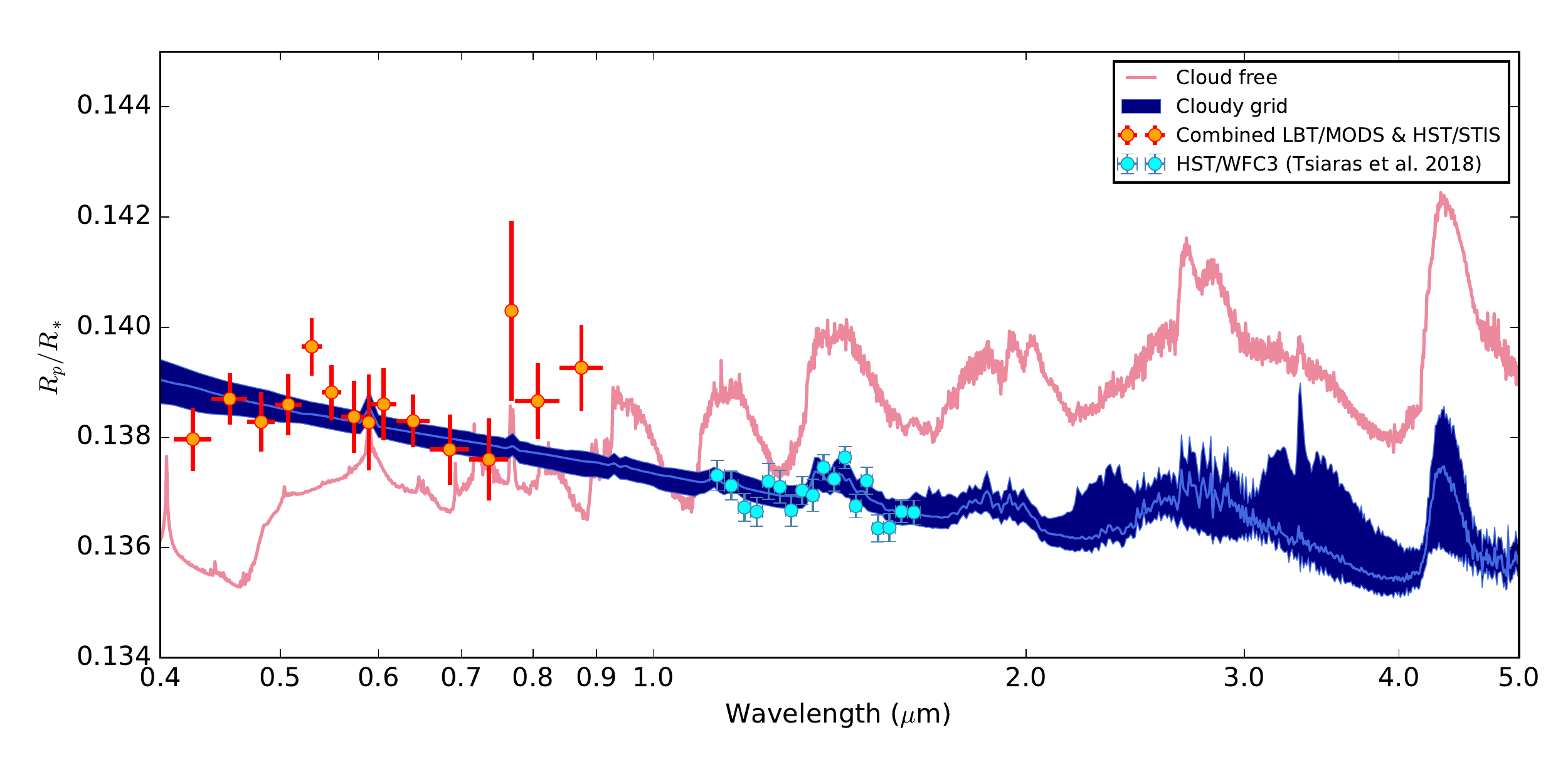}
      \caption{Combined spectrum of HAT-P-12b and the best atmospheric models fitted to this spectrum. The red points are the average of our LBT/MODS spectrum and the reanalyzed \textit{HST}/STIS spectrum from A2018. The \textit{HST}/WFC3 data (blue points) are also included in the model fit. The best-fit cloud-free model with no constraint on the atmospheric temperature (red) disagrees with the observations. The blue shadow denotes the best cloudy models with 1$\mathrm{\sigma}$ confidence region. The models are self-consistently calculated with \textit{petitCODE}. The retrieved parameters are shown in Table \ref{Retrieved_tbl}.
      }
         \label{data_model}
   \end{figure*}

In order to constrain the atmospheric properties, we combined the LBT/MODS spectrum and the \textit{HST}/STIS spectrum from A2018. The combined spectrum is presented in Fig.~\ref{data_model}. We also included the near-infrared data from the \textit{HST}/WFC3 observations \citep{Tsiaras2018}.

We used two extensive grids of self-consistent models for irradiated planets to fit the combined transmission spectrum. One grid of models was assumed to be cloud-free and was taken from \citet{molaverdikhani2019cold}. We assumed a surface gravity of 2.79 and a host star type of K5 \citep{Hartman2009}, leaving the effective temperature ($T$\textsubscript{eff}), metallicity ([Fe/H]), and carbon-to-oxygen ratio (C/O) as the three free parameters in the cloud-free models. In this grid, the temperature ranges from 400\;K to 2600\;K, metallicity from -1.0 to 2.0, and C/O from 0.25 to 1.25; covering a wide range of possibilities.

The muted water feature between 1.2 and 1.6 $\mu$m and a relatively featureless optical spectrum indicate clouds\footnote{Here the word ``cloud'' refers to the accumulation of particles, which could be haze or clouds, or a combination of both.}. Therefore clouds were taken into account for the second grid of models. The cloudy atmospheric models were calculated using \textit{petitCODE} \citep{molliere2015model,molliere2017observing} and following cloud parameterization method by \citep{ackerman_precipitating_2001}. The treatment of vertical mixing is described in Appendix A3 of \cite{molliere2017observing}. The following reactants are included in the cloudy models: \ce{H}, \ce{H2}, \ce{He}, \ce{O}, \ce{C}, \ce{N}, \ce{Mg}, \ce{Si}, \ce{Fe}, \ce{S}, \ce{AL}, \ce{Ca}, \ce{Na}, \ce{Ni}, \ce{P}, \ce{K}, \ce{Ti}, \ce{CO}, \ce{OH}, \ce{SH}, \ce{N2}, \ce{O2}, \ce{SiO}, \ce{TiO}, \ce{SiS}, \ce{H2O}, \ce{C2}, \ce{CH}, \ce{CN}, \ce{CS}, \ce{SiC}, \ce{NH}, \ce{SiH}, \ce{NO}, \ce{SN}, \ce{SiN}, \ce{SO}, \ce{S2}, \ce{C2H}, \ce{HCN}, \ce{C2H2}, \ce{CH4}, \ce{ALH}, \ce{ALOH}, \ce{AL2O}, \ce{CaOH}, \ce{MgH}, \ce{MgOH}, \ce{PH3}, \ce{CO2}, \ce{TiO2}, \ce{Si2C}, \ce{SiO2}, \ce{FeO}, \ce{NH2}, \ce{NH3}, \ce{CH2}, \ce{CH3}, \ce{H2S}, \ce{VO}, \ce{VO2}, \ce{NaCL}, \ce{KCL}, \ce{e-}, \ce{H+}, \ce{H-}, \ce{Na+}, \ce{K+}, \ce{PH2}, \ce{P2}, \ce{PS}, \ce{PO}, \ce{P4O6}, \ce{PH}, \ce{V}, \ce{VO(c)}, \ce{VO(L)}, \ce{MgSiO3(c)}, \ce{Mg2SiO4(c)}, \ce{SiC(c)}, \ce{Fe(c)}, \ce{AL2O3(c)}, \ce{Na2S(c)}, \ce{KCL(c)}, \ce{Fe(L)}, \ce{Mg2SiO4(L)}, \ce{SiC(L)}, \ce{MgSiO3(L)}, \ce{H2O(L)}, \ce{H2O(c)}, \ce{TiO(c)}, \ce{TiO(L)}, \ce{FeO(c)}, \ce{Fe2O3(c)}, \ce{Fe2SiO4(c)}, \ce{TiO2(c)}, \ce{TiO2(L)}, \ce{H3PO4(c)},  and \ce{H3PO4(L)}, where (L) and (c) denote the liquid and solid phases, respectively. The gas-phase opacities of \ce{CH4}, \ce{H2O}, \ce{CO2}, \ce{HCN}, \ce{CO}, \ce{H2}, \ce{H2S}, \ce{NH3}, \ce{OH}, \ce{C2H2}, \ce{PH3}, \ce{Na}, \ce{and K}, and the solid-phase opacities of \ce{Mg2SiO4(c)}, \ce{KCL(c)}, and \ce{Na2S(c)} are considered in these models.

In addition to the three free parameters in the cloud-free models, the sedimentation factor ($f_\mathrm{sed}$) and the geometric standard deviation of the log-normal particle-size distribution ($\sigma_{\rm g}$), are the two cloud-related free parameters in our cloudy models. In this grid, the temperature ranges from 760\;K to 1160\;K, the metallicity from -1.0 to 2.0, C/O from 0.25 to 1.25, $f_\mathrm{sed}$ from 0.01 to 3.0, and $\sigma_{\rm g}$ from 1.05 to 2.0. For a given vertical mixing strength in the atmosphere of a planet, a higher $f_\mathrm{sed}$ means a more efficient sedimentation of cloud particles in the atmosphere. A low $\sigma_{\rm g}$ value is also an indication of monodisperse particles. Both grids are publicly available\footnote{\url{www.mpia.de/homes/karan}}.

To explore the atmospheric properties of HAT-P-12b, we applied a python implementation of a Markov chain Monte Carlo (MCMC) fitting algorithm to perform a Bayesian analysis using the \texttt{emcee} tool \citep{Mackey2013}. Our approach takes into account the statistical treatment of observational uncertainties as well as uncertainties arising from the models. It can also investigate any uncertainty underestimation through a GP, where it constructs a covariance matrix iteratively. We assumed uninformative priors to initialize the walkers.

\begin{figure*}
   \centering
   \includegraphics[width=0.8\textwidth, height=0.8\textwidth]{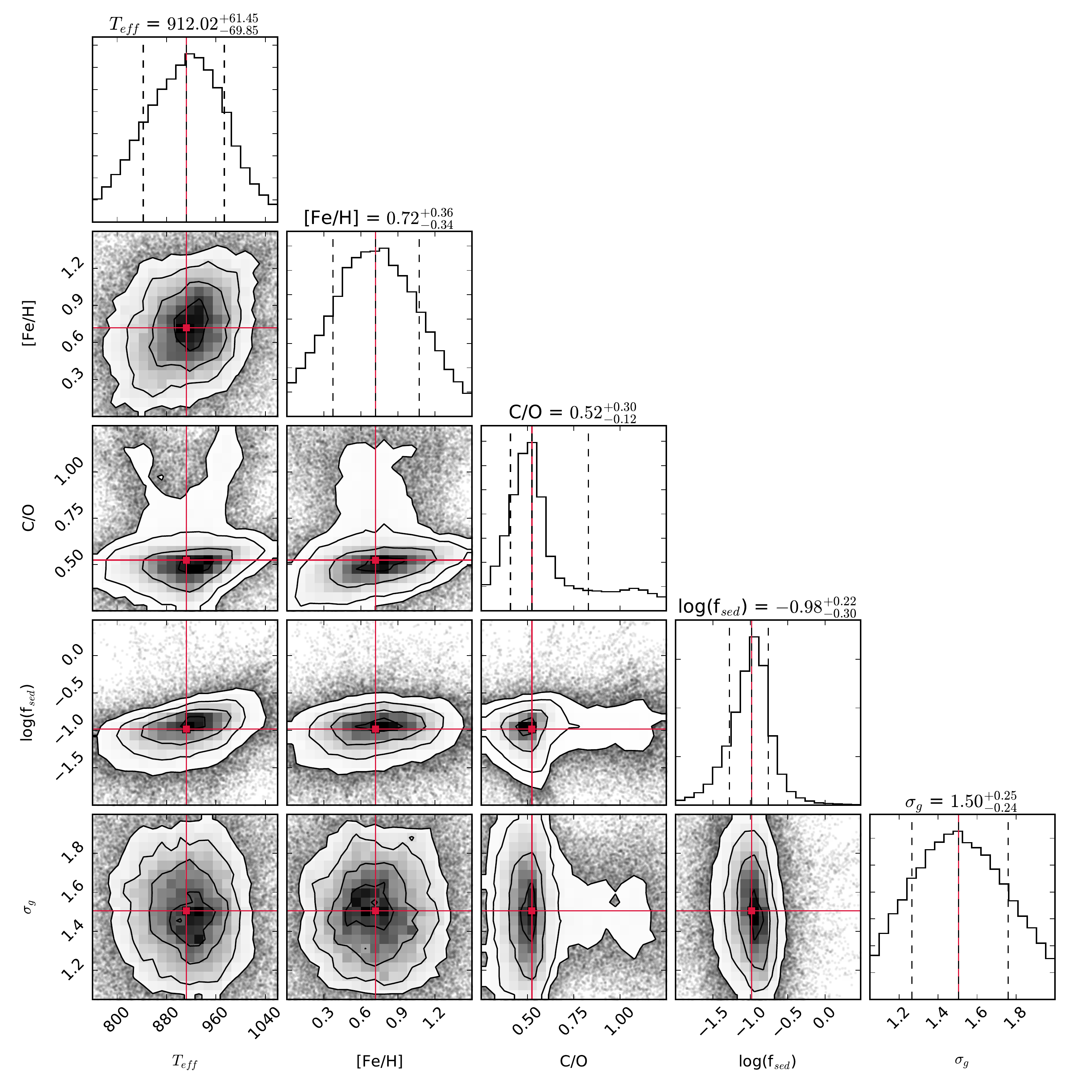}
      \caption{Correlation diagrams of the retrieved planetary atmospheric parameters.
      }
         \label{retrieve}
   \end{figure*}

The results of the MCMC analysis are shown in Fig.~\ref{data_model} (fitted spectra) and Fig.~\ref{retrieve} (corner plot of the retrieved parameters). The the best cloud-free model is shown with the red line, assuming no constraint on the temperature of the planet. The model does not represent the observed spectrum by any means. As a result, we rule out a cloud-free atmosphere for HAT-P-12b.

The cloudy grid, however, fits the combined spectrum well. We report an effective temperature of $910^{+60}_{-70}$\;K and find a supersolar metallicity of [Fe/H]$=0.72^{+0.36}_{-0.34}$ for this mildly irradiated planet. This metallicity tentatively follows the mass–metallicity relation for the Solar System planets and exoplanets \citep[e.g.,][]{Kreidberg2014b, Wakeford2017}. Our estimated carbon-to-oxygen ratio, C/O$=0.52^{+0.30}_{-0.12}$, is marginally consistent with a solar C/O$_{\rm solar}=0.54$, but a slightly subsolar C/O cannot be ruled out. Our retrieved results generally agree with the results from a recent retrieval work by \cite{Wong2020}, which used \textit{HST} and \textit{Spitzer} data of HAT-P-12b. 

As suggested by previous studies, the best-fit model is consistent with a cloudy model that maintains the cloud formation at the photospheric levels efficiently. In other words, in this case, sedimentation is likely to be inefficient in completely removing the condensates from the photosphere, leading to low $f_\mathrm{sed}$ values. For cloudy planets, a choice of $\log$($f_\mathrm{sed}$), instead of $f_\mathrm{sed}$, allows for a better exploration of very low $f_\mathrm{sed}$ region. We hence report $\log$($f_\mathrm{sed}$)$=-0.98^{+0.22}_{-0.30}$.

The relatively large uncertainties at the optical wavelengths preclude us from placing any tight constraint on the particle-size distribution parameter, although $\sigma_{\rm g}$=$1.50^{+0.25}_{-0.24}$ slightly favours monodisperse particles over polydisperse cloud particles. Further observations are required to study the scattering portion of the transmission spectrum. The results of our Bayesian analysis for the characterization of HAT-P-12b atmosphere is listed in Table \ref{Retrieved_tbl}.

We also performed the retrieval on the LBT data alone. The retrieval returns an effective temperature of $890\pm70$\;K, a metallicity of [Fe/H]$=0.77^{+0.41}_{-0.42}$, and C/O$=0.76^{+0.28}_{-0.29}$. The best-fit cloud parameters are $\log$($f_\mathrm{sed}$)$=-0.31^{+0.36}_{-0.57}$ and $\sigma_{\rm g}$=$1.54^{+0.26}_{-0.27}$.
These values are similar to the retrieved results from the combined LBT and \textit{HST} data, but with larger uncertainties.

We explored other atmospheric scenarios, including 1) a patchy atmosphere (i.e., a linear combination of cloud-free and cloudy models with the same metallicity and C/O, but allowing for different effective temperatures), 2) a cloudy atmosphere with different models accounting for dawn and dusk limbs, and 3) fitting the cloudy models with two offsets to account for stitching the LBT$+$\textit{HST}/STIS data at $\lambda$<1\;$\mu m$ to the \textit{HST}/WFC3 data at $\lambda$>1\;$\mu m$. However, introducing these complexities into the models did not improve our best fit in a statistically meaningful way. 

We only tested the stitching with two constant offsets. There might be wavelength-dependent effects when data points obtained from different observations are stitched, for example, different stellar activity levels during the observations. However, because the long-term photometry suggests that the star is not very active, the activity effect should be trivial for HAT-P-12b. Another factor is the different planetary orbital parameters used to fit the light curves obtained from different observations. 
We note that the orbital parameters ($i$ and  $a/R_*$) obtained from the \textit{HST}/WFC3 light curves in \cite{Tsiaras2018} are similar and consistent with ours and those in A2018 (Table \ref{Tab-compare}).

The best-fit model favors a cloudy atmosphere that produces a mild slope due to the scattering of cloud particles. However, we emphasize that the spectrum is still relatively flat. The slope in the best-fit model is weaker than the Rayleigh-scattering slope of a cloud-free atmosphere with a similar effective temperature. 

%

%

%
\begin{table}
\caption{Retrieved planetary atmospheric parameters from fitting self-consistent models to the combined HAT-P-12b data.}             
\label{Retrieved_tbl}      
\centering                          
\begin{threeparttable}  
\begin{tabular}{l c}        
\hline\hline                 
Planetary parameter     &        Retrieved value \\     
        \hline                       
            &   \\
        Effective temperature ($T$\textsubscript{eff})  & $910^{+60}_{-70}$\;K \\[5pt]                 
                Metallicity ([Fe/H])            & $0.72^{+0.36}_{-0.34}$ \\[5pt]
        Carbon-to-oxygen ratio (C/O)            & $0.52^{+0.30}_{-0.12}$ \\[5pt]
        Sedimentation factor ($\log$($f_\mathrm{sed}$)) & $-0.98^{+0.22}_{-0.30}$ \\[5pt]
        Width of the particle size distribution ($\sigma_{\rm g}$)                                & $1.50^{+0.25}_{-0.24}$ \\[5pt]
\hline                                   
\end{tabular}
\end{threeparttable}  
\end{table}

\begin{table}
\caption{Transit parameters of HAT-P-12b derived from this work and the other two publications.}             
\label{Tab-compare}      
\centering                          
\begin{tabular}{l c c}        
\hline\hline                 
References      &        $a/R_*$         &      $i$ [degree] \\      
        \hline                     
&   \\
\cite{Alexoudi2018} & $11.68\pm0.12$ & $88.83\pm0.19$  \\[5pt]          
\cite{Tsiaras2018} & $11.67_{-0.05}^{+0.06}$ & $88.89_{-0.08}^{+0.11}$  \\[5pt]  
This work & $11.61_{-0.15}^{+0.13}$ & $88.80_{-0.25}^{+0.31}$ \\[5pt]  
\hline                                   
\end{tabular}
\end{table}

\section{Conclusions}
We observed one transit of HAT-P-12b with the MODS multi-object spectrograph mounted on the LBT. We used the binocular and dual-channel mode of the instrument and obtained two independent sets of the planetary transmission spectrum covering $\sim$ 0.4--0.9 $\mathrm{\mu}$m. The obtained transmission spectrum is relatively flat in the visible, and there is no evidence for Na or K absorption features.  This result is inconsistent with the \textit{HST} transmission spectrum in \cite{Sing2016}, which shows a strong Rayleigh-scattering slope and a potassium feature. However,  \cite{Alexoudi2018} re-analyzed the \textit{HST} data with updated planetary orbital parameters, and they obtained a relatively flat transmission spectrum with a tentative potassium feature. Our LBT result is consistent with the reanalyzed \textit{HST} spectrum. We further compared the narrow-band transmission spectra around the K wavelengths between the LBT and \textit{HST} observations and found that the tentative potassium feature in the \textit{HST} observation might be the result of statistical fluctuations.
Therefore we conclude that the planet has a cloudy atmosphere without significant Na or K absorption features. 

We built an extensive grid of self-consistent cloudy models to fit the observed transmission spectrum. We used a combined spectrum of our LBT data and the \textit{HST} data. The fit result has a small cloud sedimentation factor, which suggests the presence of high-altitude clouds in the planetary atmosphere. 
Future observations with instruments such as the Mid-Infrared Instrument of the \textit{James Webb Space Telescope} will likely enhance our understanding of the cloud properties of this inflated sub-Saturn-mass planet.

The spectrophotometric light curves obtained from the LBT/MODS observation have precisions similar to those of the \textit{HST}/STIS observations, demonstrating that the MODS spectrograph is a powerful instrument for transmission spectroscopy studies. The capabilities of covering a wide wavelength range with a single exposure and acquiring two sets of independent spectra simultaneously (i.e., MODS1 and MODS2) are unique advantages of the MODS spectrograph.

\begin{acknowledgements}
We are grateful to the anonymous referee for their helpful comments.
We would like to thank Xanthippi Alexoudi for providing the re-analyzed \textit{HST} data and Monika Lendl, Yifan Zhou, and Jochen Heidt for useful discussions.
F.Y. acknowledges the support of the DFG priority program SPP 1992 "Exploring the Diversity of Extrasolar Planets (RE 1664/16-1)". T.H. acknowledges support from the European Research Council under the Horizon 2020 Framework Program via the ERC Advanced Grant Origins 83 24 28. A.J.\ acknowledges support from FONDECYT project 1171208, and by the Ministry for the Economy, Development, and Tourism's Programa Iniciativa Cient\'{i}fica Milenio through grant IC\,120009, awarded to the Millennium Institute of Astrophysics (MAS).
L.C. recognizes funding from Deutsches Zentrum f\"ur Luft- und Raumfahrt DLR\,-\,50OR1804. 
This material is based upon work supported by the National Aeronautics and Space Administration under Agreement No. NNX15AD94G for the program Earths in Other Solar Systems. 
This paper used data obtained with the MODS spectrographs built with
funding from NSF grant AST-9987045 and the NSF Telescope System
Instrumentation Program (TSIP), with additional funds from the Ohio
Board of Regents and the Ohio State University Office of Research.
The LBT is an international collaboration among institutions in the United States, Italy and Germany. LBT Corporation partners are: LBT Beteiligungsgesellschaft, Germany, representing the Max-Planck Society, The Leibniz Institute for Astrophysics Potsdam, and Heidelberg University; The University of Arizona on behalf of the Arizona Board of Regents; Istituto Nazionale di Astrofisica, Italy; The Ohio State University, and The Research Corporation, on behalf of The University of Notre Dame, University of Minnesota and University of Virginia. 
B.V.R. acknowledges support from the Heising-Simons Foundation.
P.M. acknowledges support from the European Research Council under the European Union's Horizon 2020 research and innovation programme under grant agreement No. 832428. G.C. acknowledges the support by the B-type Strategic Priority Program of the Chinese Academy of Sciences (Grant No. XDB41000000) and the Natural Science Foundation of Jiangsu Province (Grant No. BK20190110). 

\end{acknowledgements}

\bibliographystyle{aa} 

\bibliography{HATP12b-refers}

\end{document}